\documentclass[twocolumn]{article}
\usepackage{graphicx,amsmath,amssymb,psfrag,epsfig,multirow,color,comment}
\usepackage{natbib}
\usepackage{authblk}

\begin{document}

\title{Characterizing the dynamics of rubella relative to measles:\\the role of stochasticity}
\author[1,2]{Ganna Rozhnova\footnote{corresponding author: ganna.rozhnova@manchester.ac.uk}}
\author[3]{C. Jessica E. Metcalf}
\author[4,5]{Bryan T. Grenfell}
\affil[1]{Theoretical Physics Division, School of Physics and Astronomy,
University of Manchester, Manchester M13 9PL, United Kingdom}
\affil[2]{Centro de F{\'\i}sica da Mat\'eria Condensada and 
Departamento de F{\'\i}sica, Faculdade de Ci{\^e}ncias da Universidade de 
Lisboa, P-1649-003 Lisboa Codex, Portugal}
\affil[3]{Department of Zoology, Oxford University, Oxford, UK}
\affil[4]{Department of Ecology and Evolutionary Biology, Eno Hall, Princeton University, Princeton, USA}
\affil[5]{Fogarty International Center, National Institute of Health, Bethesda, Maryland, USA}
\date{}
\maketitle

\begin{abstract}
Rubella is a completely immunizing and mild infection in children. Understanding its behavior is of considerable public health importance because of Congenital Rubella Syndrome, which results from infection with rubella during early pregnancy and may entail a variety of birth defects. The recurrent dynamics of rubella are relatively poorly resolved, and appear to show considerable diversity globally. Here, we investigate the behavior of a stochastic seasonally forced susceptible-infected-recovered model to characterize the determinants of these dynamics and illustrate patterns by comparison with measles. We perform a systematic analysis of spectra of stochastic fluctuations around stable attractors of the corresponding deterministic model and compare them with spectra from full stochastic simulations in large populations. This approach allows us to quantify the effects of demographic stochasticity and to give a coherent picture of measles and rubella dynamics, explaining essential differences in the recurrent patterns exhibited by these diseases. We discuss the implications of our findings in the context of vaccination and changing birth rates as well as the persistence of these two childhood infections.
\end{abstract}

\section{Introduction}

\begin{figure*}
\centering
\includegraphics[trim=0cm 0cm 0cm 0cm, clip=true, width=0.69\textwidth]{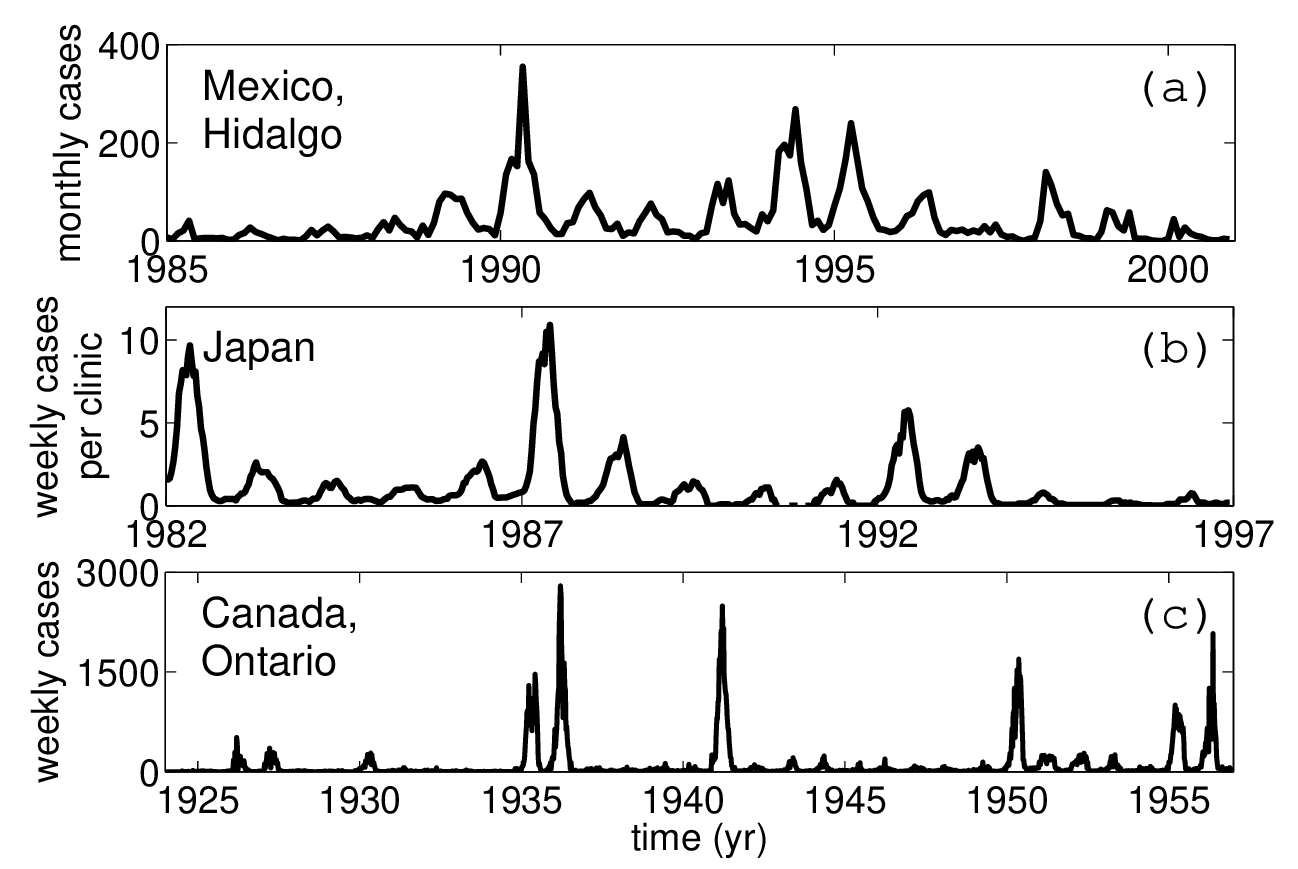}
\includegraphics[trim=0cm 0cm 0cm 0cm, clip=true, width=0.3\textwidth]{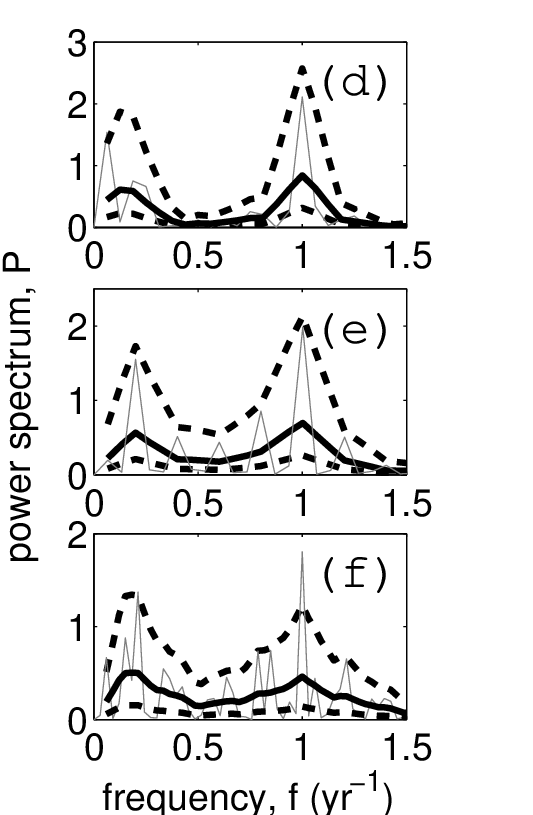}
{\caption{(Color online) Time series of the case reports of rubella and the corresponding spectrum in (a), (d) the Hidalgo district, Mexico; (b), (e) Japan and (c), (f) the province Ontario, Canada. To resolve low-frequency periodicities these time series include short intervals of vaccination (years 1998-2001 for Mexico and 1989-1992 for Japan). Before the spectrum was taken, each series was normalized, setting the mean to zero and the variance to unity. The smooth spectrum (thick black lines) was obtained from the raw spectrum (thin gray lines) using two passes of a 3-point moving average of the spectral ordinates. The dashed black lines are 90\% confidence limits on the smooth spectrum. The confidence intervals represent the uncertainly in the observations. The method of computation of the spectra and confidence limits is described in detail in Chapter 4 of Ref. \citep{diggle:1990}.}}
\label{fig0}
\end{figure*}

Rubella is a completely immunizing, directly transmitted infection, generally presenting as a mild, and potentially even asymptomatic childhood disease \citep{best:2007}. As a result, rubella tends to be under-reported, and its recurrent dynamics are fairly poorly characterized. Nevertheless, since infection during early pregnancy may cause spontaneous abortion or Congenital Rubella Syndrome (CRS), which may entail a variety of birth defects \citep{cooper:1967}, understanding the dynamics of rubella is of considerable public health importance. Dynamical features of rubella may alter the CRS burden via their effects on the average age of infection. Episodic dynamics may increase the average age of infection, as the intervals between larger outbreaks provide the opportunity for individuals to age into later age classes \citep{ferrari:2010, heij:98}. Likewise, local extinction dynamics can allow individuals to remain susceptible as they age into their childbearing years \citep{metcalf:2011a, metcalf:2012}, resulting in the potential for a considerable CRS burden once rubella is re-introduced.

Empirically, rubella seems to be linked to either i) annual dynamics, as in Mexico \citep{metcalf:2011}, Peru \citep{metcalf:2011a}, or parts of Africa \citep{goodson:2011,metcalf:southafrica}; ii) spiky dynamics, as in Canada \citep{bauch:2003} and iii) some hint at multi-annual regularity, as in Japan \citep{terada:2003}, England and Wales \citep{anderson:1991}, and various European countries \citep{edmunds:2000}. In Figure 1 we show three time series that represent the range of observed rubella dynamics. Spectral analyses of time series are particularly useful for understanding temporal patterns exhibited by different data \citep{pristley:1981,diggle:1990}. The characteristic feature of rubella spectra are an annual peak at 1 year and a multi-annual peak at 5-6 years exhibited by all data in Figure 1. Rubella also seems to experience regular fade-outs \citep{metcalf:2011}, which is of great epidemiological importance, particularly in the context of increasing global control efforts. The propensity for stochastic extinction is characterised by the critical community size (CCS), below which the infection tends to go extinct in epidemic troughs. Analyses of dynamics in Mexico and Peru suggest a CCS of over $10^6$ for rubella \citep{metcalf:2011, metcalf:2011a}.

Measles provides a natural comparison for rubella, as it is perhaps the most extensively studied of the childhood infections, and its dynamics are very well understood \citep{rohani:99,bharti:2010,bjorn:02,conlanBirth:2007,earn:00,ferrari:2010,ferrari:08,finkenstadt:2002,finkenstadt:98,grenfell:02,grenfell:01,keelinggrenfell:02,Mantilla:2010}. Before the start of vaccination in England and Wales, both biennial dynamics (e.g. in London) and annual dynamics (e.g. in Liverpool) were observed \citep{keelinggrenfell:02,earn:00,bjorn:02,grenfell:02}. The underlying driver of this variability has been identified as differences in birth rate, combined with annual seasonality in transmission driven by school term times \citep{schenzleD:84,bjorn:02,keelinggrenfell:02,conlanBirth:2007}. In sub-Saharan Africa, chaotic dynamics have been shown to result from a very high birth rate, combined with extreme seasonal forcing \citep{ferrari:08}. Both highly irregular dynamics \cite{rohani:99,earn:00} (e.g. following vaccination in England and Wales) and triennial dynamics \citep{london:73} (e.g. in Baltimore between 1928 and 1935) have also been reported. The spectral analyses of measles data exhibiting the described dynamics can be found in the now standard textbook \citep{anderson:1991}. The CCS of measles is rather smaller than that of rubella, estimated by Bartlett \citep{bartlett:57} to be between $2.5\times10^5-3\times10^5$ for England and Wales.

The two key ingredients underlying models of childhood diseases like rubella and measles are i) seasonality in transmission due to schooling patterns and ii) demographic stochasticity arising from the discrete nature of population \citep{schenzleD:84, durrett:1994, keelinggrenfell:02, nguyen:08}. Although various approaches have been used to understand the dynamics of rubella \citep{keeling:01, bauch:2003, anderson:86, boo:87}, most of the analyses have been essentially deterministic. Keeling et. al. \citep{keeling:01} considered a term-time forced susceptible-infected-recovered (SIR) model and compared its dynamics to rubella data in Copenhagen. From this, they concluded that the dynamics of rubella may result from switching between two cyclic attractors (annual and multi-annual limit cycles) of the deterministic model. Although the deterministic analysis they present is comprehensive, there is only a limited amount of evidence to suggest that the switching will occur in contexts that include demographic stochasticity. In particular, in this work \citep{keeling:01} stochasticity was introduced into the model as multiplicative noise of arbitrary amplitude instead of using, for instance, standard stochastic simulations based on the Gillespie algorithm (for unforced models) \citep{gillespie:76} and its extensions (for seasonally forced models) \citep{anderson:07}. Such simulations produce exact realisations of the stochastic process, whose full dynamics is given by the solution of the master equation as described in Section \ref{theoreticalanalysis}. For large populations the master equation is approximated by the deterministic model with additive noise \citep{rozhnova:10,blackb:10}.

Bauch et. al. \citep{bauch:2003} studied a term-time forced susceptible-exposed-infected-recovered (SEIR) model and showed that frequencies obtained from the linear perturbation analysis of the deterministic model are in good agreement with positions of the peaks in spectra of data records of various childhood infections. The application of this approach to rubella data for Canada predicted two distinct peaks at 1 and 5.1 years, close to what we see in Figure 1 (f). With the exception of Ref. \citep{bauch:2003}, where stochastic simulations for Canada parameter values were also performed, there has been no work on rubella using a fully stochastic approach dealing with demographic stochasticity.

Here, we use this approach to characterize different rubella dynamics and illustrate patterns by comparison with measles. To this end, we perform the theoretical analysis of spectra of stochastic fluctuations around stable attractors of a seasonally forced deterministic SIR model and compare them with spectra obtained from full stochastic simulations based on a modification \citep{anderson:07} of the algorithm by Gillespie \citep{gillespie:76}. The mathematical techniques employed in this work have been developed for ecological and epidemiological models \citep{mckane:05, rozhnova:10, rozhnova:11} and applied to model temporal patterns of measles and pertussis \citep{rozhnova:12, blackb:10, black:10}. The picture that emerges to explain rubella dynamics is close to that proposed in \citep{bauch:2003} but goes beyond it because our analysis allows us to obtain the full structure of a spectrum (as opposed to the deterministic analysis of \citep{bauch:2003} where only frequencies of the spectral peaks could be predicted). By introducing key spectral statistics (described below) we systematically investigate how the dominant period, amplitude and coherence of stochastic fluctuations change across a broad range of epidemiological parameters. We then discuss the implications of our analysis in the context of changing birth rates and vaccination levels, as well as their implications for the persistence of measles and rubella.

\section{Methods}

\subsection{Model}
The individual-based stochastic model we explore in this paper follows a standard seasonally forced SIR structure \citep{keeling:08, anderson:1991}. At any time $t$, it consists of a discrete population of constant size $N$ divided into compartments of susceptible, $S(t)$, infected, $I(t)$, and recovered, $R(t)$, individuals. Susceptible individuals become infected (and infectious) at a frequency dependent rate $\beta(t)I(t)/N$, where $\beta(t)$ is a seasonally varying transmission rate. For childhood diseases, $\beta(t)$ captures an increase in the number of contacts between school children during terms with respect to holidays \citep{schenzleD:84}. Both term-time and sinusoidal forcing have been used to model these changes \cite{keeling:01,blackb:10,bauch:2003,earn:00,kuznetsov:94,stone:12,rozhnova:10}. Previous studies \cite{bauch:2003,earn:00} have shown that the form of forcing is not crucial for the essential dynamic structure (the bifurcation diagram) of the sinusoidally and term-time forced models if the seasonal forcing amplitude is adjusted appropriately. We therefore focus on a sinusoidally forced $\beta(t)=\beta_0 (1+\epsilon\cos 2\pi t)$, where $\beta_0$ is the average transmission rate and $\epsilon$ is the amplitude of seasonal forcing, and confirm later that the dynamic temporal patterns observed in simulations of the term-time forced model are similar. Infected individuals recover at constant rate $\nu$ ($1/\nu$ is the average infectious period). As is common in the mathematical epidemiology literature \citep{keeling:08, anderson:1991}, we restrict our attention to the case when birth and death rates $\mu$ ($1/\mu$ is the average lifetime) are equal, and thus the total population size $N$ is constant. This allows us to reduce the number of independent variables to two and define the state of the system as $\sigma=\{S(t),I(t)\}$. From $\beta_0$, $\nu$ and $\mu$ we can express one of the most important epidemiological parameters \citep{keeling:08, anderson:1991}, the basic reproductive ratio $R_0=\beta_0/(\nu+\mu)$. $R_0$ is the average number of secondary cases caused by one infectious introduced into a fully susceptible population; $R_0$ will be used throughout the text.

\subsection{Theoretical Analysis}
\label{theoreticalanalysis}
Two main approaches can be used to investigate the dynamics of the stochastic model formulated above. An analytical approach starts from the formulation of the model as a master equation for the probability of finding the system in state $\sigma$ with $S(t)$ susceptibles and $I(t)$ infectives at time $t$ \citep{kampen:1981, risken:1996, gardiner:2003, keelingross:2008}. Much understanding about the stochastic dynamics relevant for recurrent epidemics can be gained if this equation is expanded in powers of $1/\sqrt{N}$ \citep{kampen:1981}. An extensive discussion of this approach has already been given at length in the literature in the context of epidemic models, and we refer the reader to \citep{alonso:07, blackb:10, rozhnova:10} for formal details. Here we describe only the aspects which are important for this paper. In essence, the method involves the substitutions $S(t)=N\overline{s(t)}+\sqrt{N}x_S(t)$ and $I(t)=N\overline{i(t)}+\sqrt{N}x_I(t)$ in the master equation which can be then expanded to obtain two systems of equations \citep{kampen:1981}. At the leading order, the expansion gives rise to a set of ordinary differential equations for the {\it mean} variables, i.e. the fractions (densities) of susceptible and infected individuals, $\overline{s(t)}$ and $\overline{i(t)}$. These equations are the same as the standard deterministic SIR model with sinusoidal forcing (see e.g. \citep{keelinggrenfell:02}). At next-to-leading order it gives rise to a set of stochastic differential equations for {\it fluctuations} of susceptible, $x_S(t)$, and infected, $x_I(t)$, individuals about the mean behavior given by the deterministic model \citep{kampen:1981}. From these equations we are able to analytically calculate power spectra of fluctuations for susceptibles, $P_S(f)$, and infectives, $P_I(f)$, as functions of frequency $f$. We are interested in the endemic behavior of the model so the spectra correspond to the fluctuations about {\it stable attractors} of the deterministic model which for $\epsilon=0$ and $\epsilon>0$ are the endemic {\it fixed point} \citep{alonso:07} and stable {\it limit cycles} with a period that is an integer multiple of a year \citep{blackb:10, rozhnova:10}, respectively. Further technical details relating to analytical calculations are given in the Supplementary Material. Throughout the text we will use the words {\it theoretical} and {\it analytical} interchangeably to refer to spectra computed as explained in this section.
\begin{figure}
\centering
\includegraphics[trim=0cm 0cm 0cm 0cm, clip=true, width=0.5\textwidth]{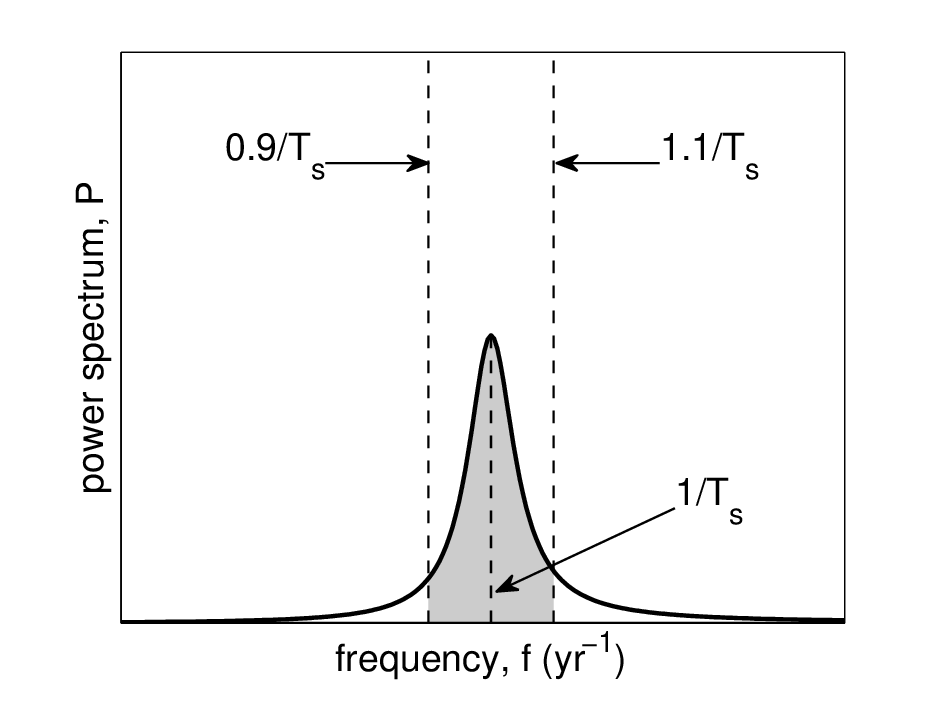}
{\caption{Schematic plot of a power spectrum of stochastic fluctuations for infectives, $P(f)$. The quantities used in the comparative analysis of different spectra are the dominant period (the inverse of the main peak's frequency), amplification (total area under the power spectrum curve) and coherence (ratio of the shaded area to the total area).}}
\label{fig1}
\end{figure}

In our analysis we will focus on a spectrum $P_I(f)$, which for simplicity will be denoted as $P(f)$, and its three characteristics, namely {\it dominant period}, {\it amplification} and {\it coherence} \citep{alonso:07, simoes:08}, see Figure 2. We define the dominant period of stochastic fluctuations as inverse frequency of the maximum of the highest stochastic peak. We also compute the total spectral power which equals the area under a power spectrum curve. This quantity defines the ability of the system to sustain oscillations of all frequencies and shall be referred to as the amplification of stochastic fluctuations. Finally, the coherence is defined as the ratio of spectral power lying within 10\% from the dominant period and the total spectral power. It serves to measure how well-structured oscillations about the dominant period are.

As a rule, a theoretical spectrum of unforced epidemic models ($\epsilon=0$) has one peak \citep{alonso:07, rozhnova:12}, while it can have several peaks of different amplitudes for $\epsilon>0$ \citep{blackb:10, rozhnova:10}. Away from bifurcation points of the deterministic model, one of them is usually much higher than the others. We are not aware of any work assessing the relevance of secondary peaks to recurrent epidemic behavior seen in real data. The highest peak, however, has been shown to be important in understanding the inter-epidemic periods observed in time series of pertussis and measles \citep{blackb:10, black:10, rozhnova:12}, and is therefore used in the definition of a spectrum's characteristics in this paper.

\subsection{Simulations}
\label{simulations}
We simulate the model using an extension of Gillespie's algorithm \citep{anderson:07, gillespie:76} which produces stochastic trajectories for $\{S(t),I(t)\}$ in continuous time. These are processed further to compute numerical spectra and test them against the theoretical prediction for $P(f)$. The simulation length is 500 years and the first 50 years are discarded. In numerical work a time series for fluctuations $x_I(t)$ is obtained as $x_I(t)=\left[I(t)-N\overline{i(t)}\right]/\sqrt{N}$, where $\overline{i(t)}$ is the fraction of infectives averaged over many realizations of the model. From $x_I(t)$ we compute a spectrum $P(f)$ using the discrete Fourier transform. For $\epsilon>0$ we also present a spectrum of the entire `signal' (scaled by population size $N$), $I(t)$, which will be referred to as a {\it full} spectrum. By definition $P(f)$ includes only stochastic peaks, while the full spectrum includes both deterministic peaks corresponding to a limit cycle and stochastic peaks corresponding to fluctuations about it. For either of these spectra we will use the words {\it simulated} and {\it numerical} interchangeably to emphasize that they were computed using the method described in this section. For each set of parameters, 250 simulations are recorded and all final spectra are averaged over those where no extinctions occurred during 500 years. The initial conditions for susceptibles and infectives are chosen from $S(0)=round( s_cN ) + U(0,30)$ and $I(0)=round( i_cN ) + U(0,30)$, where $round(\cdot)$ is rounding to the nearest integer, $U$ is the uniform distribution and $s_c$ and $i_c$ is the fixed point (for $\epsilon=0$) or a random point on the limit cycle (for $\epsilon>0$). The random number generators used in the Gillespie algorithm are initialized with unrepeated seeds which guarantees that the simulated stochastic trajectories are all different. We have also checked that with this choice of initial conditions all simulations converge to a stationary state within 50 years (transient period) or die out and so are not taken into consideration.

Both the theoretical analysis described in Section \ref{theoreticalanalysis} and the numerical analysis based on simulations described in Section \ref{simulations} are suitable for the investigation of temporal patterns in large populations, such as those corresponding to the time series in Figure 1, and both have advantages and limitations. The simulations can be quite easily implemented but progressively become computationally intensive as the population size, $N$, increases. In a systematic study as we will perform here, the numerical analysis would become very time-consuming for populations larger than 1 million individuals. It is for those populations, the approximate analytical spectrum, computed from the expansion in the inverse population size, is instead expected to predict the dynamics very well (the better the large the population size is) \citep{blackb:10}. In the case when the seasonality is absent, the analytical spectrum is given by a simple formula (see the Supplementary Material) which can be readily used to compute spectral characteristics. For the seasonally forced model the spectrum can be written as an analytical formula too \citep{blackb:10,rozhnova:10}, however the calculation is more evolved and has to be done numerically because no closed expression for a limit cycle can be found. The theoretical analysis also helps to understand the mechanisms behind the dynamics such as the change in temporal patterns when approaching bifurcation points of the deterministic model, that are not always clear from a visual inspection of simulated time series. The limitations of the theory are that (a) it does not allow us to compute the spectrum of fluctuations when the deterministic model has several stable coexisting limit cycles for a given set of parameters and (b) in the vicinity of bifurcations the perfect agreement with simulations is achieved for populations larger than 1 million. Both cases will be discussed in detail in the next section.   

The data time series presented in Figure 1 correspond to large populations (Hidalgo's population size was about 2.1 million \citep{metcalf:2011} and the other time series are likely to correspond to even larger populations). In the following section we perform the systematic comparison of recurrent patterns for a large range of realistic values of the infectious period and the basic reproductive ratios. For the sake of computational speed we focus on the population of 1 million of individuals and we demonstrate how the spectral analysis can be used to predict the dynamics in even larger populations.

\begin{figure*}
\centering
\includegraphics[trim=0.6cm 0.1cm 1.8cm 0.4cm, clip=true, width=0.32\textwidth]{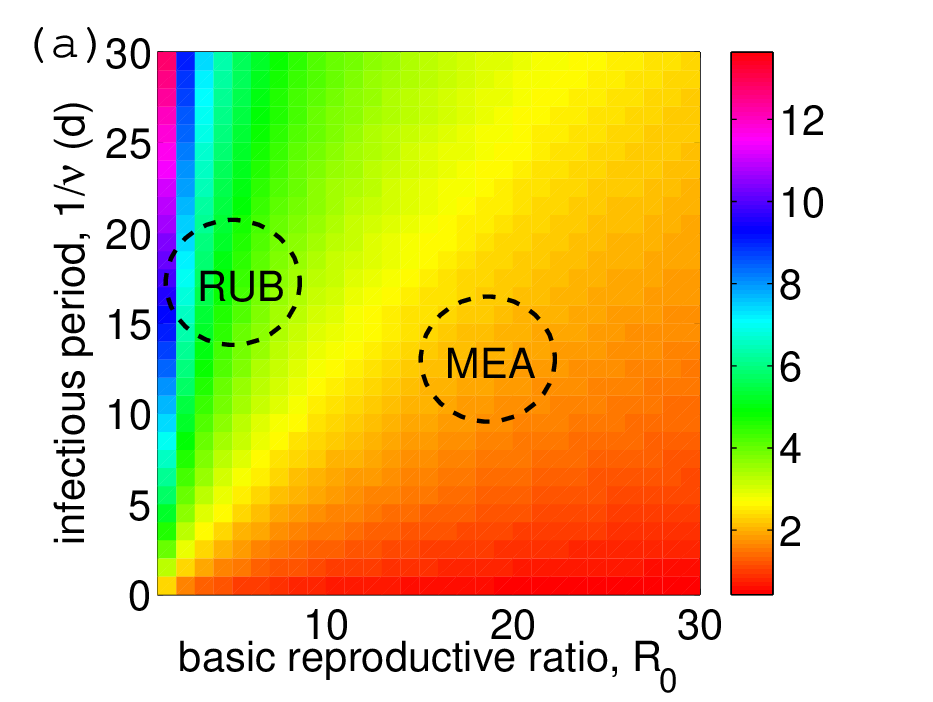}
\includegraphics[trim=0.6cm 0.1cm 1.6cm 0.4cm, clip=true, width=0.32\textwidth]{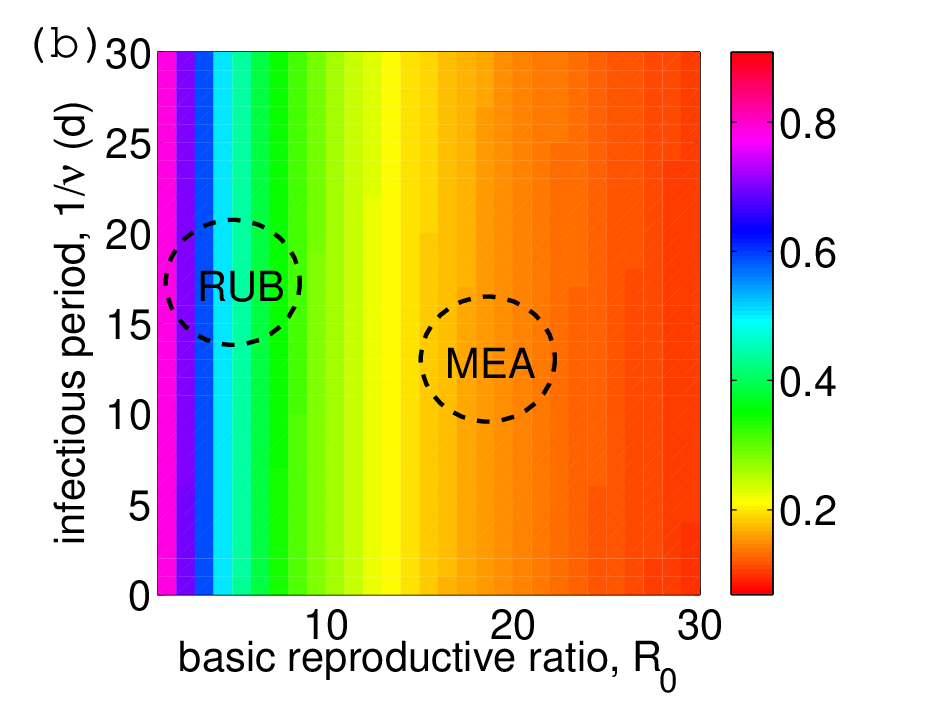}
\includegraphics[trim=0.6cm 0.1cm 1.6cm 0.4cm, clip=true, width=0.32\textwidth]{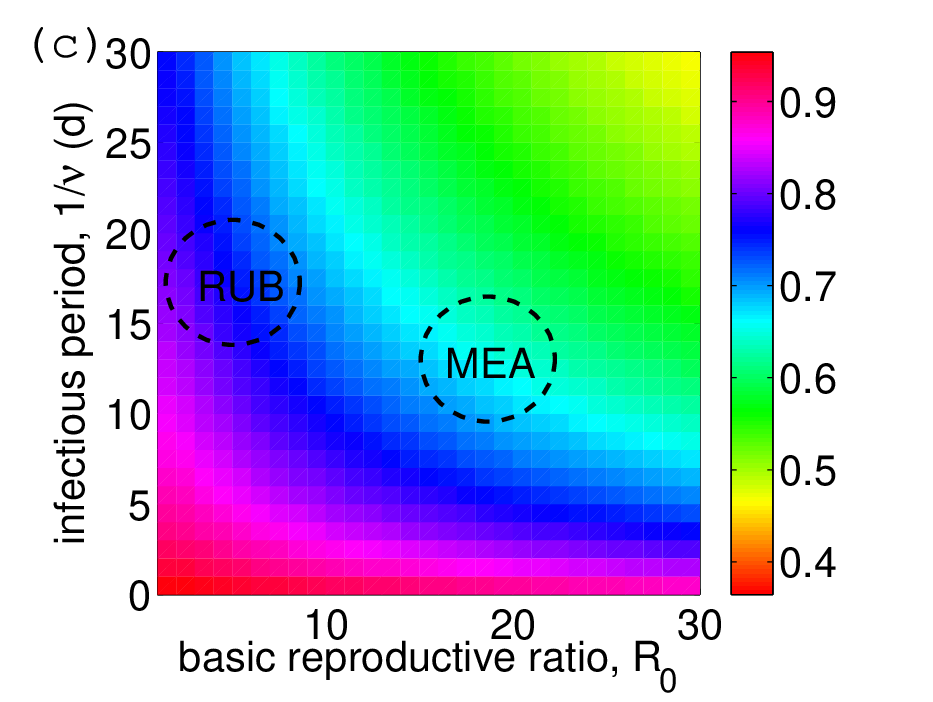}\\
\includegraphics[trim=0.6cm 0.1cm 1.8cm 0.4cm, clip=true, width=0.32\textwidth]{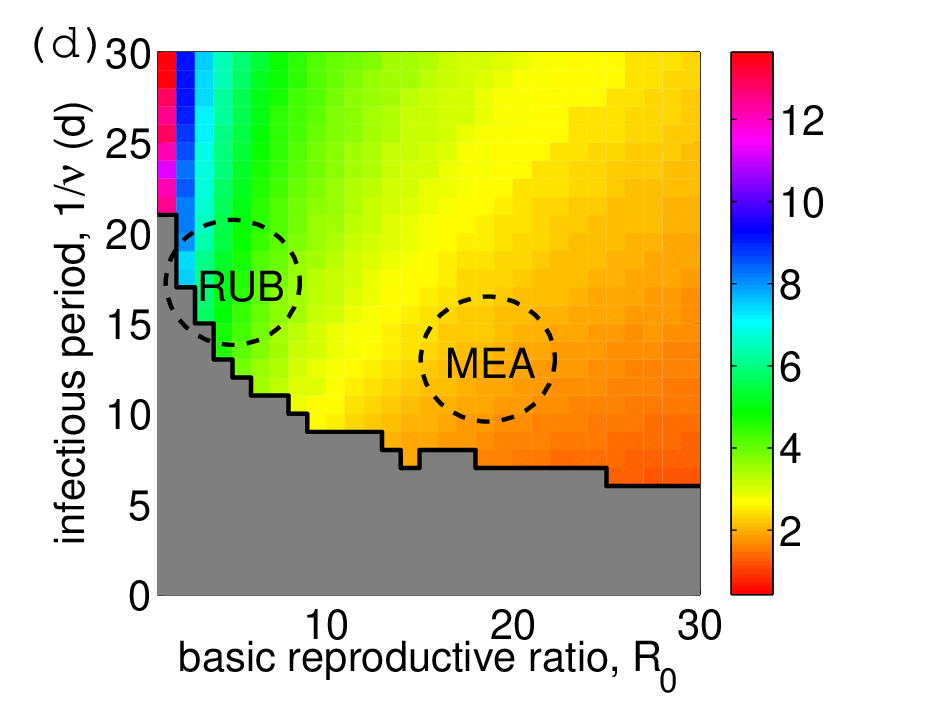}
\includegraphics[trim=0.6cm 0.1cm 1.6cm 0.4cm, clip=true, width=0.32\textwidth]{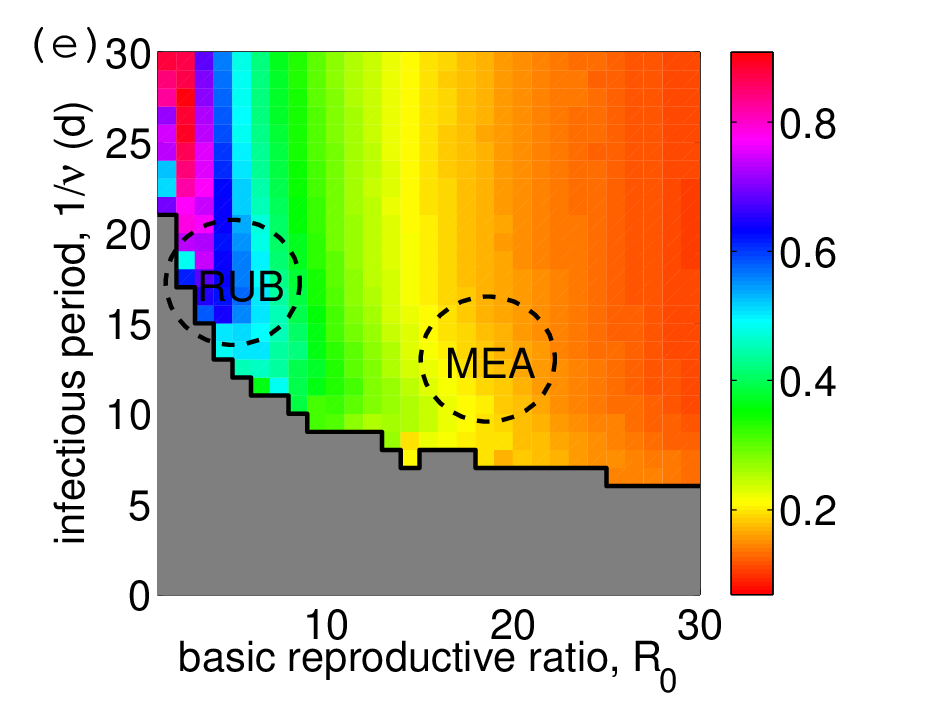}
\includegraphics[trim=0.6cm 0.1cm 1.6cm 0.4cm, clip=true, width=0.32\textwidth]{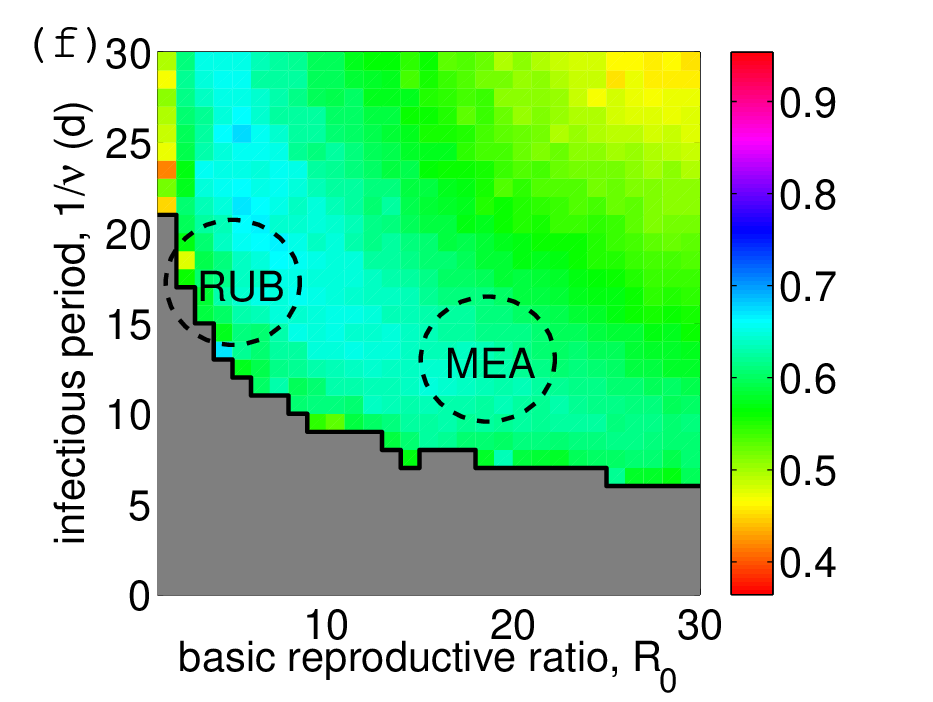}
{\caption{(Color online) Analytical (top) and simulation (bottom) results for the (a), (d) dominant period of stochastic fluctuations about the endemic fixed point, (b), (e) amplification and (c), (f) coherence. The black stairstep graph bounds the gray region where all simulations went extinct within 500 years. Approximate parameter values for measles and rubella are depicted as dashed circles. Parameters: $\epsilon=0$, $\mu=0.02$ 1/y and $N=10^6$.}}
\label{fig3}
\end{figure*}

\section{Results}

We compare the numerical and theoretical predictions for different spectra and the three measures we have conventionally chosen to characterise them. In the beginning we explore a large region of parameter space and later discuss the main findings for rubella and measles.

\subsection{No seasonal forcing: $\epsilon=0$}
\subsubsection{Theoretical and simulation results}
We first restrict our attention to the case when there is no seasonality, for which an explicit expression for the analytical spectrum can be found (see the Supplementary Material). The deterministic SIR model has only one endemic fixed point provided $R_0>1$ \citep{keeling:08, anderson:1991}. Figure 3 shows analytical and numerical results for the dominant period of stochastic fluctuations about it as well as their amplification and coherence for a range of basic reproductive ratios, $R_0$, and infectious periods, $1/\nu$. Analytical spectra can be obtained for any $R_0>1$. In practice long numerical simulations may not be feasible for all parameter combinations where $R_0>1$ because the system experiences frequent extinctions when the infectious period is short. The results are presented in Figure 3, and the domain where this happens is shown in gray. In what follows, the line separating the gray region from the rest of the parameter space will be called the extinction boundary. We would like to stress that this concept is used for our convenience and is different from the concept of the CCS. In particular, its location depends on the population size, the length of time series and the number of runs used in simulations. For populations smaller/larger than 1 million individuals and simulation length larger/smaller than 500 years the gray region would be extended/abridged and the extinction boundary would be shifted. In the region of parameter space amenable to the exploration of the long-term dynamics of the model, we observe that the structure of the spectra uncovered by the theory is clearly visible in the simulations too. Small systematic deviations between the two predictions are expected and occur close to the extinction boundary. These are due to the non-Gaussian character of the fluctuations which cannot be explained within the theory used in this study (see the Supplementary Material). To achieve the perfect agreement with the simulations within this region, the theoretical analysis would require considering the next order corrections of the order $\sqrt{N}$ to the macroscopic equations. The deviations are mainly reflected in the broadening of a spectrum and appearance of secondary peaks. As a consequence, amplification, see Figure 3 (e), is slightly increased (coherence, Figure 3 (f), is correspondingly decreased) in the simulations in the area adjacent to the gray region. In addition to these changes, an increase of the dominant period of stochastic fluctuations in the simulated spectra may be observed \citep{simoes:08,rozhnova:10}. This effect will be discussed in more detail in the next section where seasonality is included as, without seasonality, it is only barely apparent (compare Figure 3 (a) and (d)).

\subsubsection{Implications for rubella and measles dynamics}
The structure discovered in Figure 3 allows us to derive an initial picture of the dynamics of rubella relative to measles. Based on estimates of parameters typical of these diseases for the pre-vaccination era \citep{anderson:1991, keeling:01, keelinggrenfell:02, bauch:2003, bjorn:02, metcalf:2011} we superimpose their approximate locations in all panels of Figure 3. The rubella estimates are for Mexico and Canada, and the measles estimates are for England and Wales. For rubella the infectious period, $1/\nu$, is about 18 days and $R_0$ ranges from 3.4 to 9.5 in Mexico (Figure 1 (a)) \citep{metcalf:2011}. $R_0$ in the Canadian province Ontario ranges from 4.6 to 6.5 where the lower bound is the estimate for the years shown in Figure 1 (c) \citep{bauch:2003}. For measles we have taken the most frequently used values for large cities (e.g. London) in England and Wales before vaccination, $1/\nu$ about 2 weeks and $R_0$ around 18. 

The results so far ignore the seasonality of transmission rate and so are insufficient to explain the patterns of measles in which it plays a pivotal role \citep{bjorn:02, ferrari:2011, keelinggrenfell:02}. However, they have important implications for understanding the dynamics of rubella. As we shall confirm shortly, for $\epsilon>0$ the spectrum of stochastic fluctuations for this disease is close in form to that obtained for the unforced model which correctly predicts a dominant period associated to the stochastic peak of about 5-6 years (Figure 3 (a) and (d)) as seen from the comparison with the left peaks in the data spectra (Figure 1). This period is similar to the natural period of small amplitude perturbations from the endemic fixed point recovered in the purely deterministic setting \citep{keeling:01}. 

Our analysis of the stochastic model allows us to quantify other features of fluctuations using amplification and coherence, see Figure 3 (b)-(c) and (e)-(f). For rubella, the amplification is large indicating that the epidemic patterns of the unforced model represent high amplitude oscillations. High coherence suggests that only a few of the frequencies involved in the stochastic fluctuations account for most of the variance of time series. This peculiar type of dynamics sets rubella close to the extinction boundary. Large coherent multi-annual epidemics with troughs deeper than in the region with higher $R_0$ cause regular extinctions. In the next section, we discuss how these descriptions of the stochastic dynamics of rubella are changed in the presence of seasonality and compare it with the dynamics of measles.

\begin{figure*}
\centering
\includegraphics[trim=0cm 0cm 0cm 0cm, clip=true, width=0.48\textwidth]{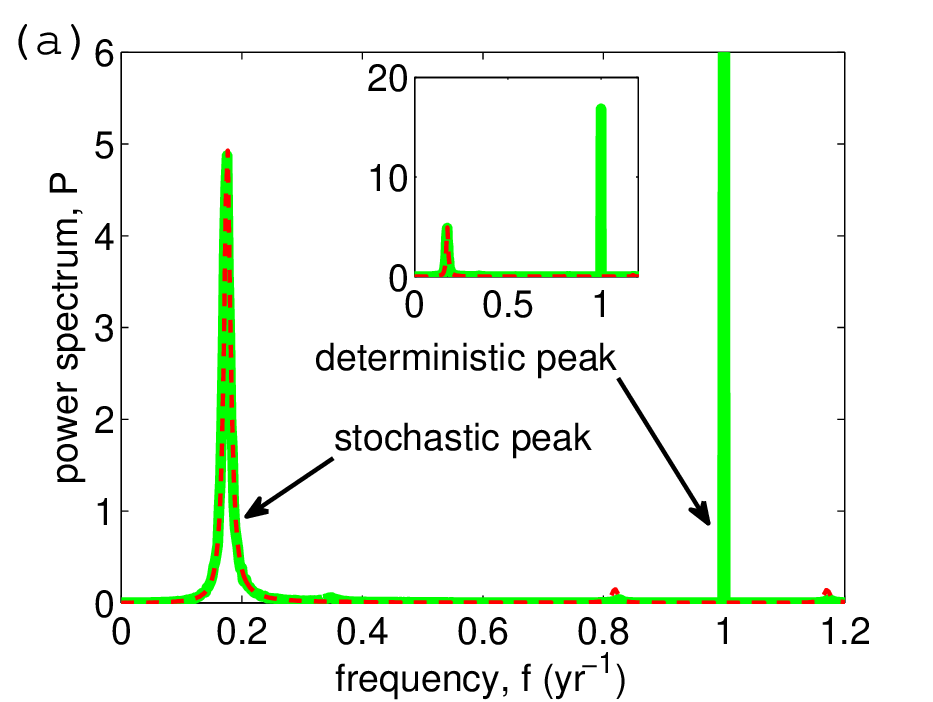}
\includegraphics[trim=0cm 0cm 0cm 0cm, clip=true, width=0.48\textwidth]{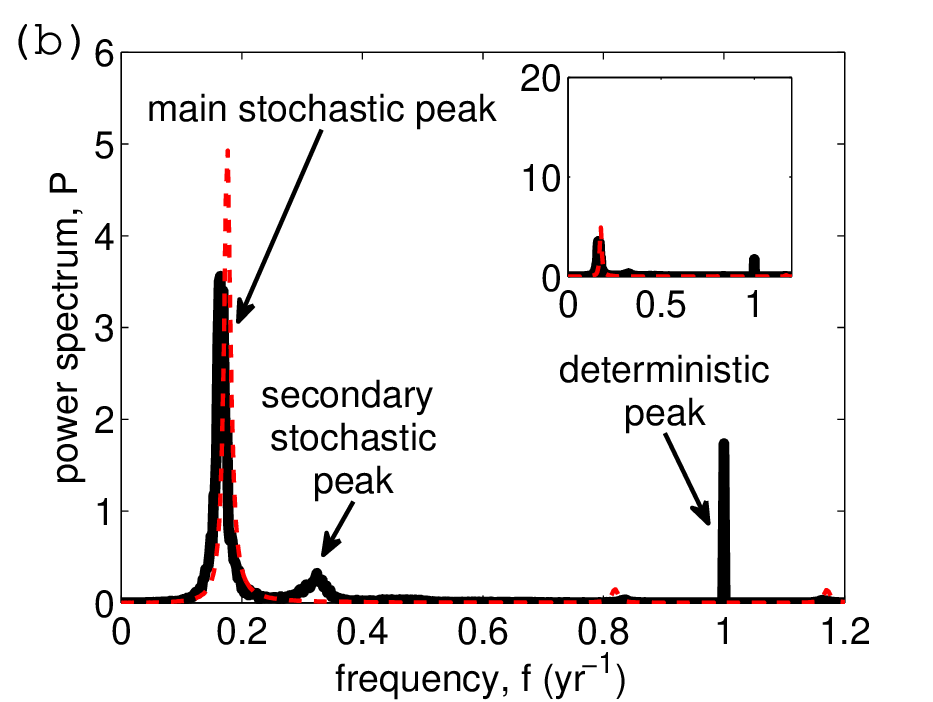}
{\caption{(Color online) Analytical spectra of stochastic fluctuations around an annual cycle (dashed red lines) and full numerical spectra (solid green and black lines) for the seasonally forced model. The dashed red line is practically indistinguishable from the solid green line, deviating only for values corresponding to frequencies of around 1 year. For clarity in the comparison the peak at 1 year is cropped in the main panel of (a) while the entire spectra can be seen on the same scale in the insets. Parameters: $\epsilon=0.05$, $\mu=0.02$ 1/y, $R_0=4$, $1/\nu=18$ d, (a) $N=10^7$ (green) and (b) $N=10^6$ (black).}}
\label{fig5}
\end{figure*}

\subsection{Seasonal forcing: $\epsilon>0$}
\subsubsection{Theoretical and simulation results}
For $\epsilon>0$ the spectra are associated with stochastic fluctuations about stable attractors of the deterministic model, i.e. stable limit cycles of a period in multiples of a year \citep{rozhnova:10, blackb:10}. The seasonally forced deterministic SIR model has a complex bifurcation diagram with regimes where multiple limit cycles may coexist \citep{kuznetsov:94, stone:12}. For high birth rates and high seasonality regions corresponding to chaotic dynamics are found \citep{ferrari:08}. Across most of the range of parameter space we explore here, either annual or biennial limit cycles are present. We performed the theoretical analysis for these attractors for different parameters and found that the agreement between the theory and simulations is in general excellent. Nevertheless, small discrepancies are again expected if a limit cycle is not sufficiently stable and/or a population is small. In particular, this happens near the extinction boundary, and is therefore relevant for rubella. 


\begin{figure*}
\centering
\includegraphics[trim=0.6cm 0.1cm 1.8cm 0.4cm, clip=true, width=0.32\textwidth]{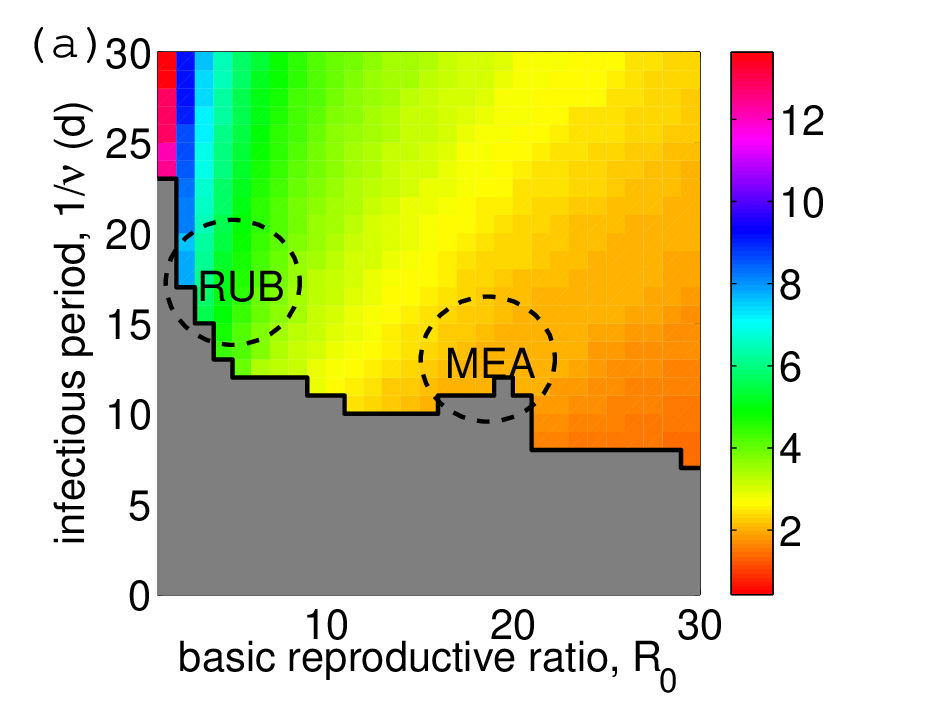}
\includegraphics[trim=0.6cm 0.1cm 1.6cm 0.4cm, clip=true, width=0.32\textwidth]{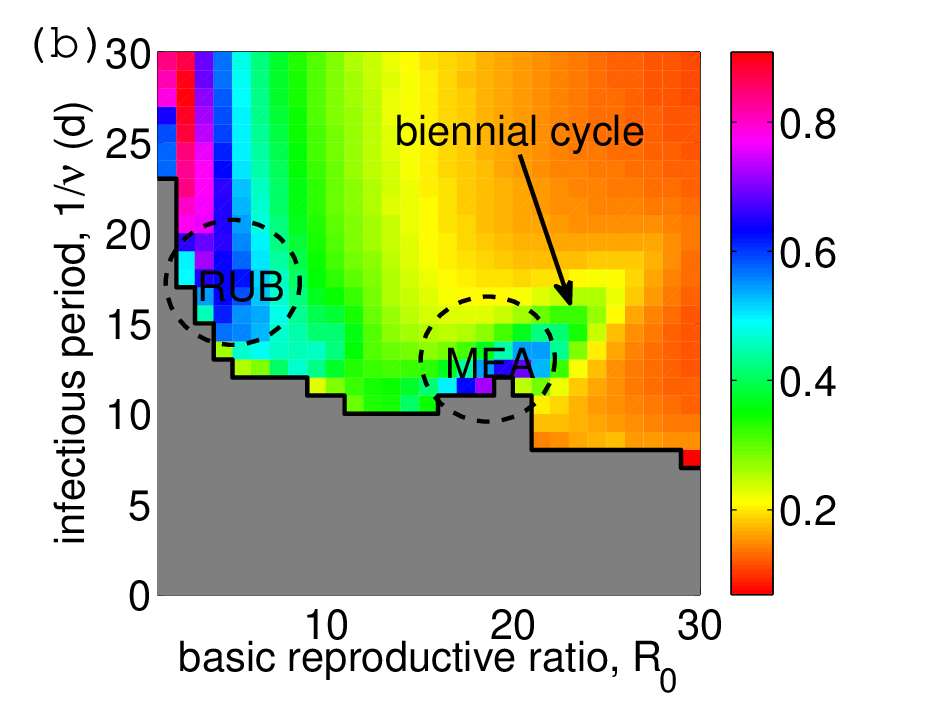}
\includegraphics[trim=0.6cm 0.1cm 1.6cm 0.4cm, clip=true, width=0.32\textwidth]{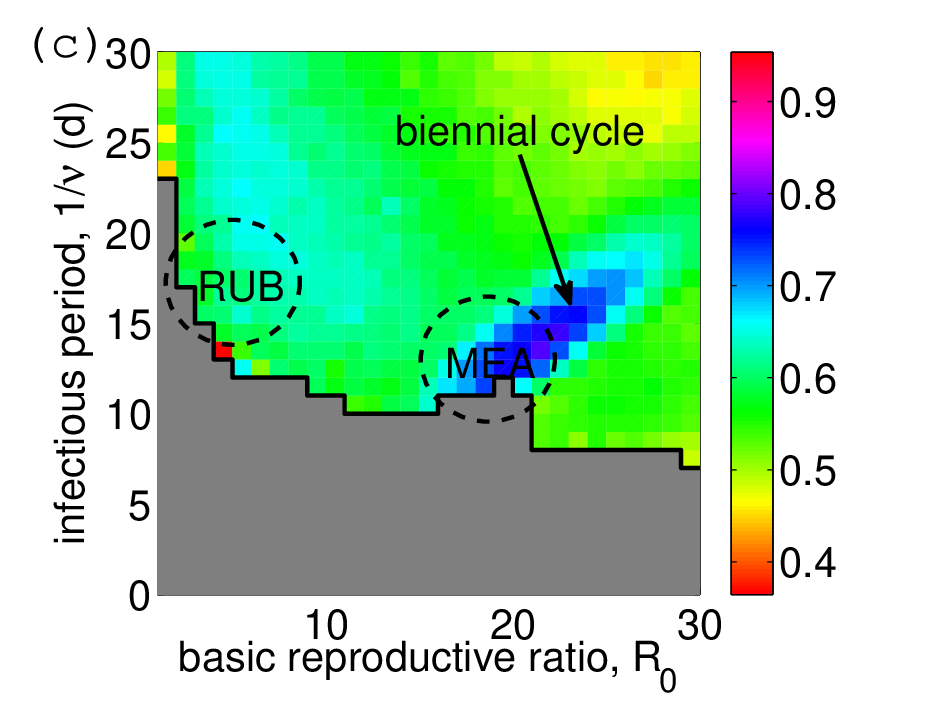}\\
\includegraphics[trim=0.6cm 0.1cm 1.8cm 0.4cm, clip=true, width=0.32\textwidth]{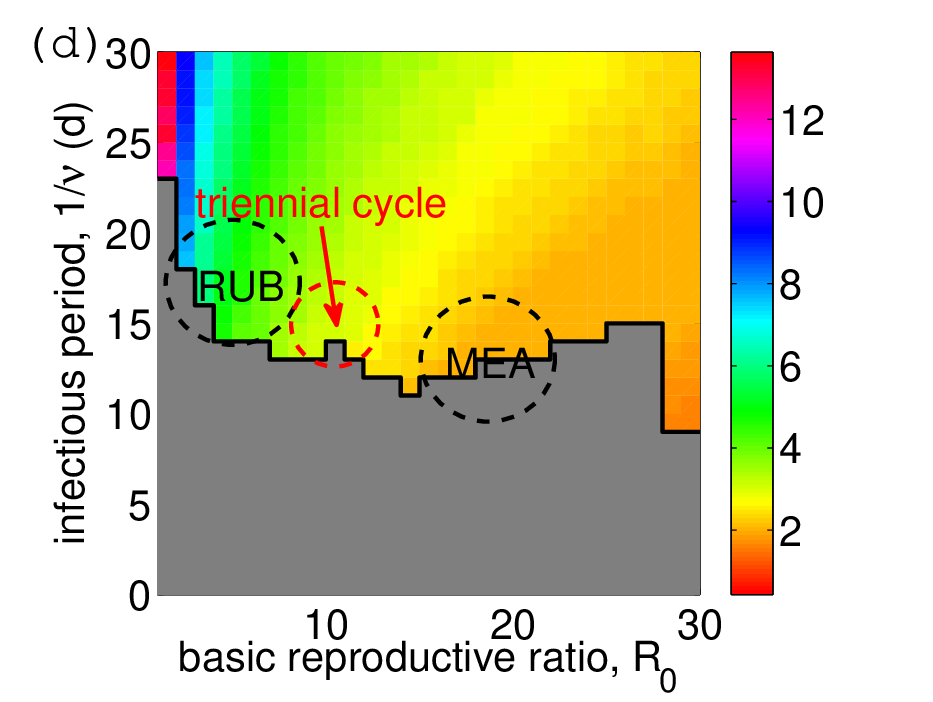}
\includegraphics[trim=0.6cm 0.1cm 1.6cm 0.4cm, clip=true, width=0.32\textwidth]{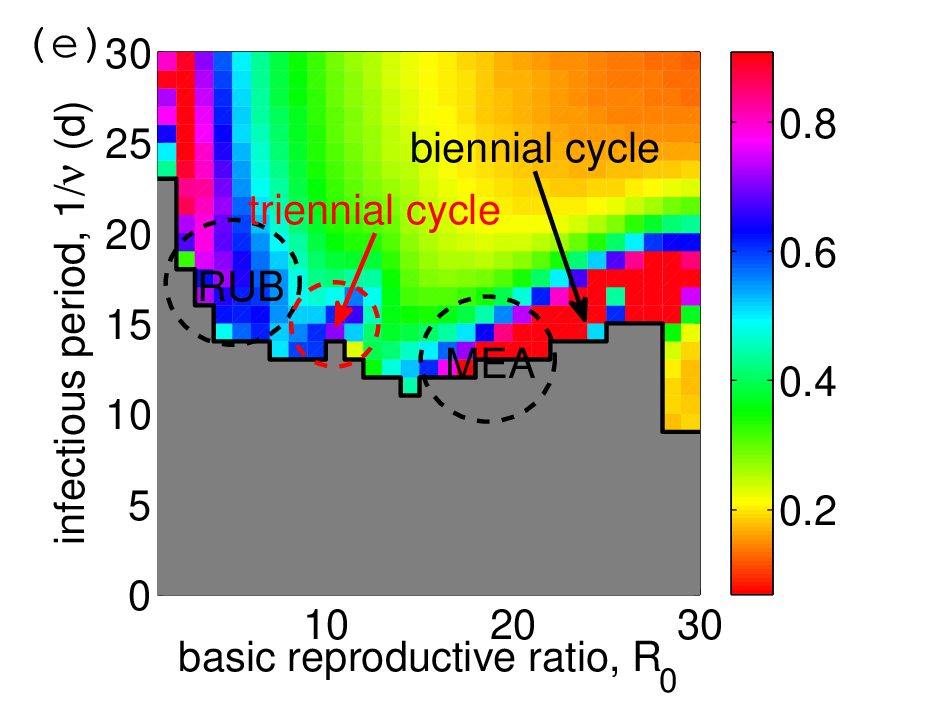}
\includegraphics[trim=0.6cm 0.1cm 1.6cm 0.4cm, clip=true, width=0.32\textwidth]{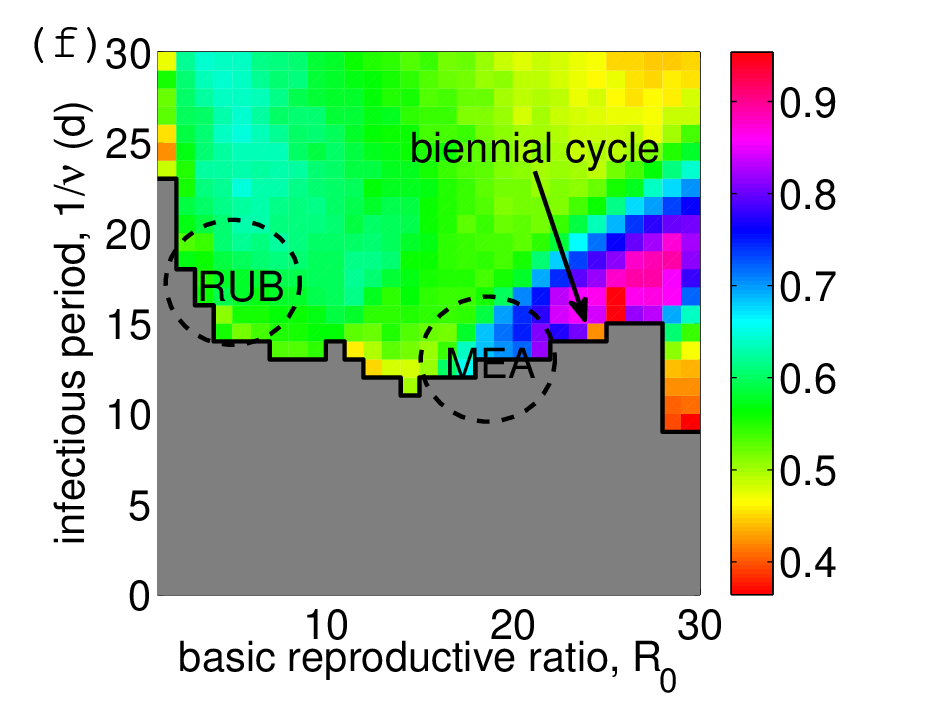}\\
\includegraphics[trim=0.6cm 0.1cm 1.8cm 0.4cm, clip=true, width=0.32\textwidth]{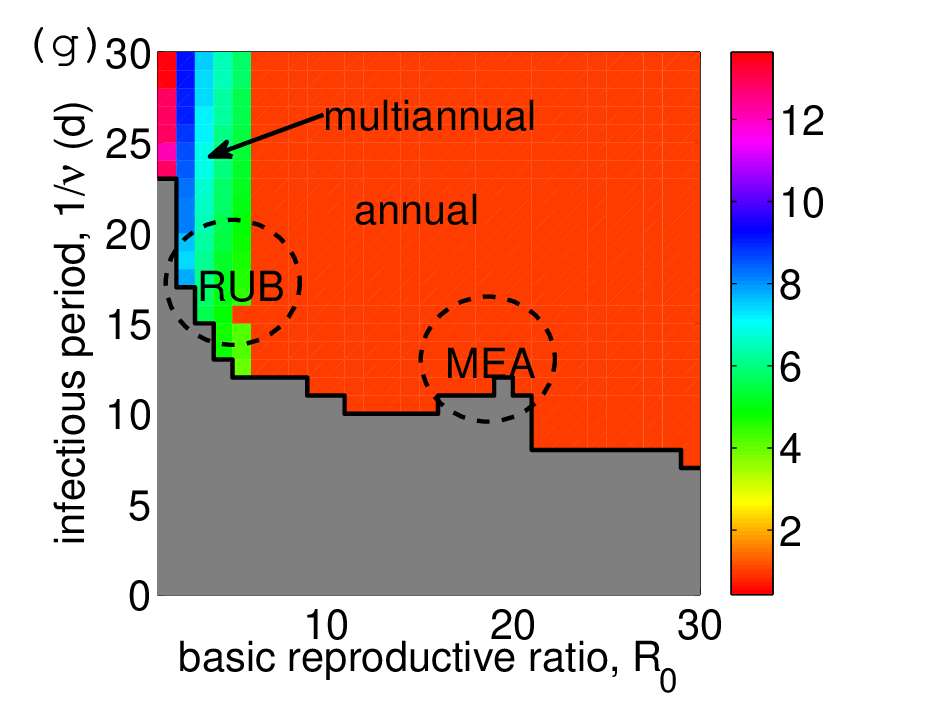}
\includegraphics[trim=0.6cm 0.1cm 1.6cm 0.4cm, clip=true, width=0.32\textwidth]{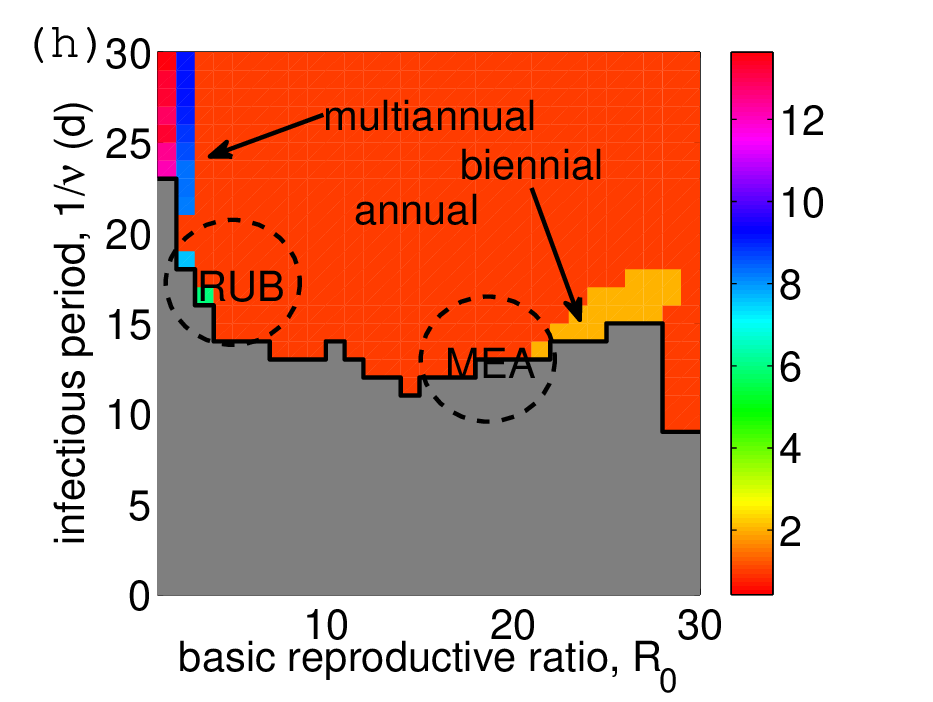}
{\caption{(Color online) Simulation results for the (a), (d) dominant period of stochastic fluctuations, (b), (e) amplification, (c), (f) coherence and (g), (h) dominant period in the full spectrum. The seasonality is twice larger in the second row and panel (h) than in the first row and panel (g). The black stairstep graph bounds the gray region where all simulations went extinction within 500 years. Approximate parameter values for measles and rubella are depicted as dashed circles. Parameters: $\mu=0.02$ 1/y, $N=10^6$, (a)-(c) and (g) $\epsilon=0.05$, (d)-(f) and (h) $\epsilon=0.1$.}}
\label{fig6}
\end{figure*}

For parameters reflecting rubella, an annual limit cycle is found in the deterministic model. To illustrate the effects of population size on simulated spectra we show in Figure 4 an analytical spectrum about this attractor (dashed red line) and full numerical spectra for $N=10^7$ (solid green line) and $N=10^6$ (solid black line). For the larger population size, see Figure 4 (a), the simulated spectrum exhibits two types of peaks (solid green line). There is a dominant annual peak (corresponding to the deterministic annual cycle) and a subdominant broad multi-annual peak (corresponding to stochastic fluctuations about it). The latter is indistinguishable from the theoretical spectrum (dashed red and solid green lines coincide, deviating only for values corresponding to frequencies of around 1 year). The derivation of an approximate theoretical spectrum (dashed red curve) from the expansion in powers of $1/\sqrt{N}$ suggests that the stochastic fluctuations, $x_I(t)$, are Gaussian ($\langle x_I(t)\rangle=0$), see the Supplementary Material. This allows us to represent the full spectrum of $I(t)=N\overline{i(t)}+\sqrt{N}x_I(t)$ as the sum of two parts: a deterministic part that scales as $N^2$ and a stochastic part that scales as $(\sqrt{N})^2$. The full spectra in our analysis are normalized (divided) by $N$ as we mention in Section 2.3. Therefore, for large populations where there is a perfect agreement between the analytical and simulated spectra as in Figure 4 (a), the amplitude of the deterministic peak is proportional to $N$ and the amplitude of the stochastic peak is independent of $N$.

This scaling is captured even for smaller populations where the fluctuations become non-Gaussian. For $N=10^6$ the peak at 1 year becomes subdominant, see the solid black line in Figure 4 (b). This indicates that the contribution of an annual component in the time series decreases with decreasing $N$. As for fluctuations beyond the annual component, at least two stochastic peaks at 5.8 and 2.9 years can be clearly seen (solid black line). Although the theory does not capture them in full, the agreement is still good and, more importantly, the systematic qualitative changes can be predicted. For small populations, the dominant period of fluctuations in simulations is slightly increased and their variance is distributed over a larger range of frequencies with respect to theoretical predictions \citep{rozhnova:10}. This is compatible with a general observation of the increased stochasticity and therefore much more irregular dynamics in small populations \citep{grenfell:02, lloyd:09}. We would like to point out that the discussed discrepancies are not attributed to sample size effects (the number of runs used to compute the spectrum). The latter may lead to discrepancies only for parameters at the very border with the black line (see Figure 1 in the Supplementary Material). The example we presented here was for $R_0=4$, which is 2 points away from the extinction boundary. Simulations for larger $R_0$ show smaller deviations from analytical calculations even for populations as small as $N=10^6$ (see Figure 2 in the Supplementary Material). As mentioned before we expect these results to be robust to the form of seasonal forcing. This is confirmed in Figure 3 of the Supplementary Material which shows that the simulated spectra from Figure 4 are reproduced by the term-time forced model at a 2.7 larger forcing amplitude.

Another situation which the analytical theory cannot fully account for is stochastic switching between different attractors of the deterministic model \citep{earn:00, keeling:01}. The computation of an analytical spectrum about a limit cycle requires the knowledge of its geometric orbit (see Supplementary Material). In the sinusoidally forced SIR model several stable attractors coexist in the regions of small $R_0$ and $1/\nu$ \citep{kuznetsov:94} and spectra of stochastic fluctuations about each of them can be obtained separately \citep{rozhnova:10}. The theoretical analysis, however, does not allow us to predict which of the attractors will be observed in simulations and what their relative contribution to the total stochastic dynamics is. Previous analysis of the stochastic dynamics of measles and pertussis showed that the only attractors seen in simulations of the seasonally forced SIR model (and other related models of infectious diseases) are annual and biennial cycles \citep{rozhnova:10, blackb:10, rozhnova:12}. The stochastic switching was observed to happen exclusively between these attractors and only for measles. This result is, however, of limited value to us because it is restricted to particular parameter choices, and so we cannot assume that the switching does not happen in the broader span of the parameter space.

Spectra from simulations contain complete information about the frequency distribution of oscillations and are thus helpful to identify switching between attractors through the presence of unexpected peaks. Figure 5 shows simulation results for the dominant period of stochastic fluctuations, amplification and coherence for two values of seasonality $\epsilon$ and $N=10^6$. In addition to these quantities we compute the dominant period in the full spectrum which includes both stochastic and deterministic peaks.

\begin{figure*}
\centering
\includegraphics[trim=0cm 0cm 0cm 0cm, clip=true, width=0.48\textwidth]{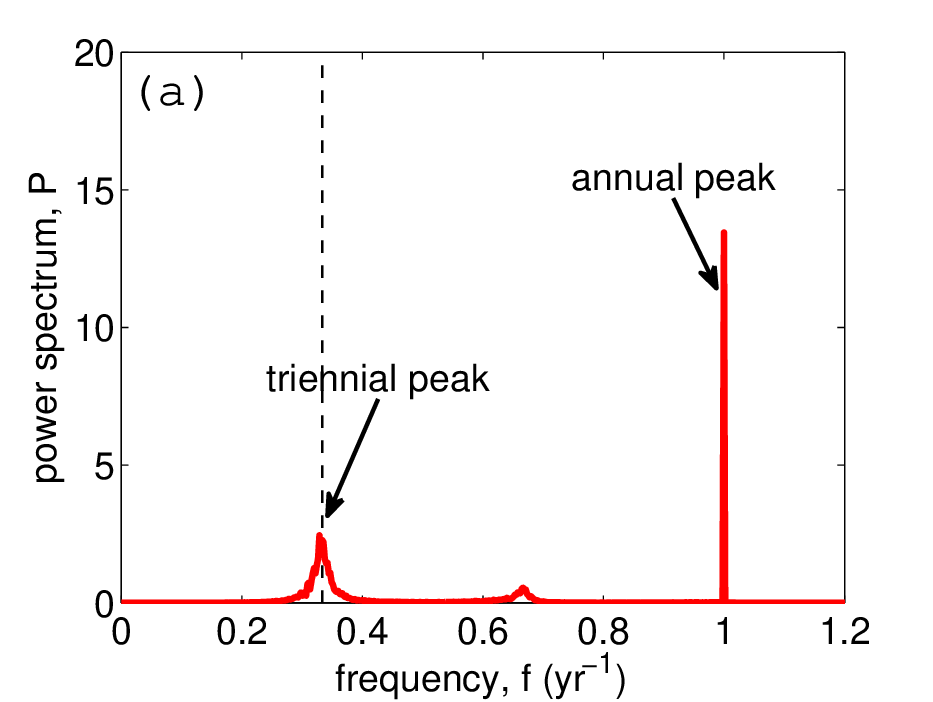}
\includegraphics[trim=0cm 0cm 0cm 0cm, clip=true, width=0.48\textwidth]{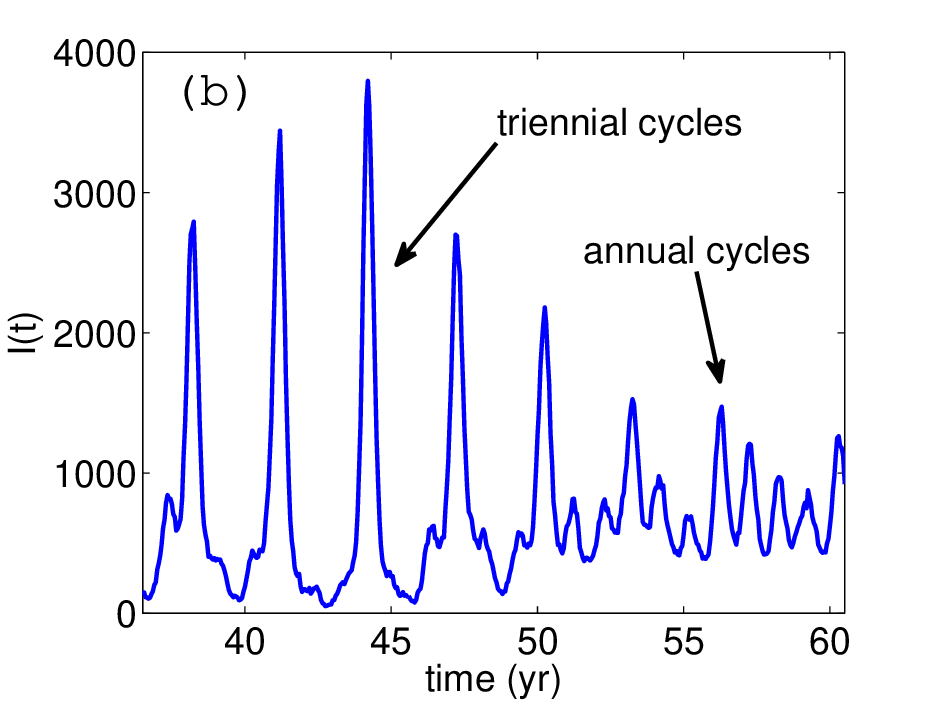}
{\caption{(Color online) (a) Full numerical spectrum and (b) typical time series corresponding to the stochastic switching between the annual and triennial cycles in the seasonally forced model. Parameters: $N=10^6$, $\mu=0.02$ 1/y, $R_0=11$, $1/\nu=15$ d, $\epsilon=0.1$. The dashed line in (a) indicates the frequency corresponding to the period 3 years.}}
\label{fig6}
\end{figure*}

To examine the effect of seasonality on stochastic fluctuations Figure 5 (a)-(f) can be directly compared with Figure 3 (a)-(f) for the unforced model. As $\epsilon$ increases the (gray) domain with frequent extinctions is extended and approaches the measles parameters. For most values of $R_0$ and $1/\nu$ we have explored (the colored region) the dynamics of the stochastic model are associated with fluctuations about only few attractors. Firstly, the biennial cycle is found inside the region of increased coherence and amplification in Figure 5 (b)-(c) and (e)-(f) which includes measles and is absent in Figure 3 (b)-(c) and (e)-(f). The dominant period in the full spectrum of time series demonstrating such a behavior is at 2 years (Figure 5 (h)). Secondly, stochastic switching between annual and triennial cycles is detected in a small region relevant for measles with low values of $R_0$, see an unexpected increase of amplification in Figure 5 (e) around $R_0=10$. The spectra here have a dominant annual peak (Figure 5 (h)) and a subdominant triennial peak (Figure 5 (d)). The amplification of oscillations associated with the latter is however much higher than what we would expect to see for fluctuations around an annual cycle (see, e.g. Figure 5 (b)). In Figure 6 we show the full spectrum and a typical time series corresponding to the switching between the annual and triennial cycles. Thirdly, in the rest of the (colored) region which includes rubella, the spectra are similar to those of the unforced model. The dynamics of the stochastic model here corresponds to fluctuations about an annual cycle and we discuss it first. 

\begin{figure*}
\centering
\includegraphics[trim=0cm 0cm 0cm 0cm, clip=true, width=0.48\textwidth]{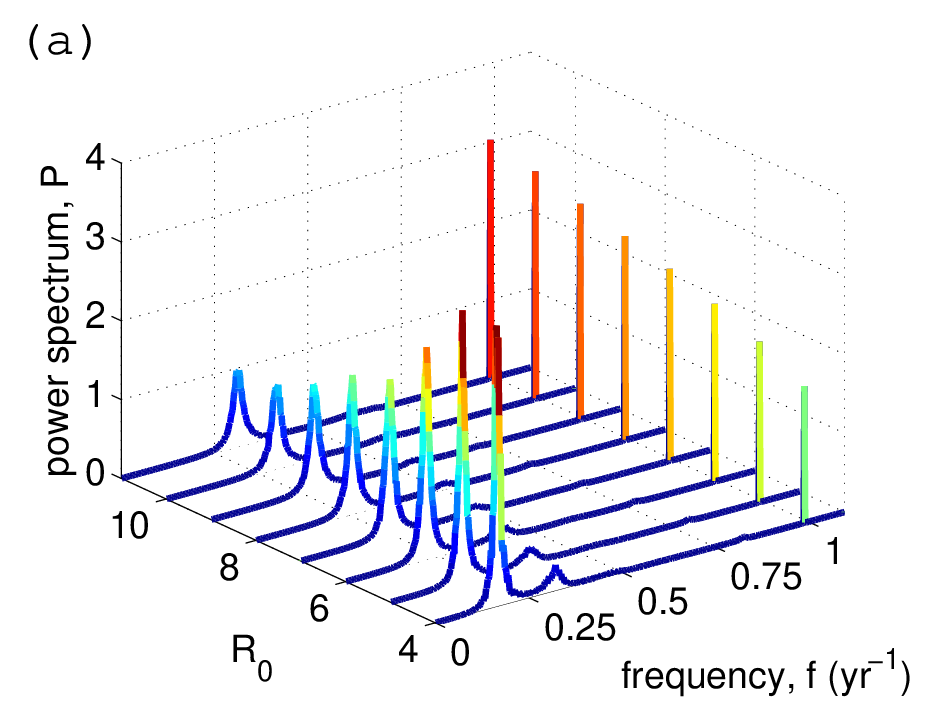}
\includegraphics[trim=0cm 0cm 0cm 0cm, clip=true, width=0.48\textwidth]{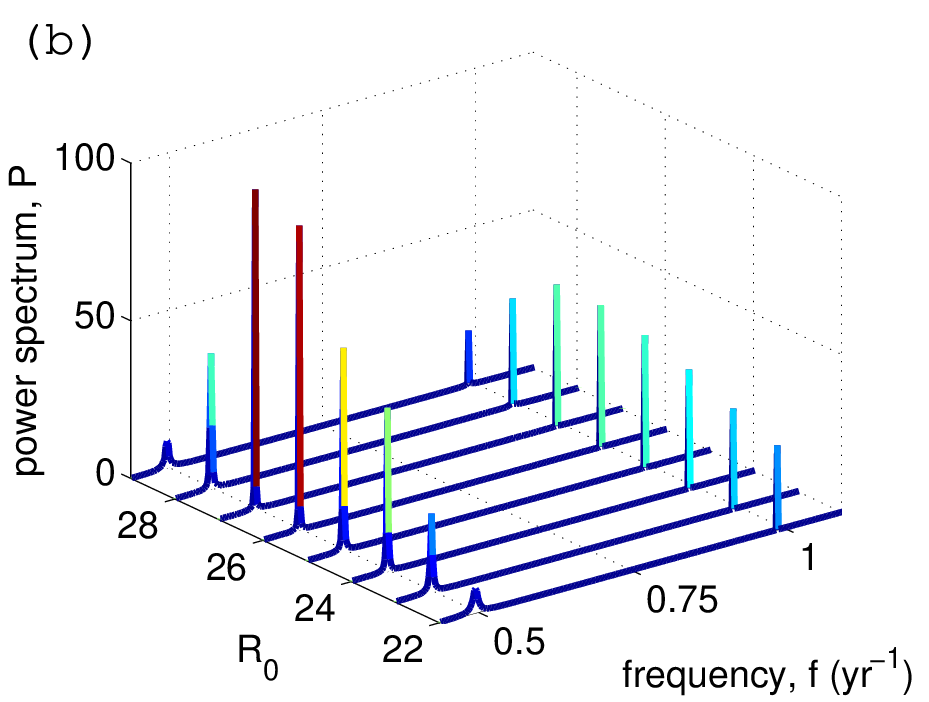}
{\caption{(Color online) Full spectra from simulated time series, corresponding to the dynamics typical of (a) rubella and (b) measles. The color intensity increases (from blue to red) with increasing power. Parameters: $\mu=0.02$ 1/y, $N=10^6$, (a) $\epsilon=0.05$, $1/\nu=18$ d, (b) $\epsilon=0.1$, $1/\nu=16$ d.}}
\label{fig7}
\end{figure*}

\subsubsection{Implications for rubella and measles dynamics}

From Figure 5 (a) and (d) we see that for relatively small basic reproductive ratios, typical of rubella, the seasonality does not affect the dominant period of stochastic fluctuations which continues to be centered at about 5-6 years. The amplification (coherence) is only slightly increased (decreased) as $\epsilon$ increases (Figure 5 (b)-(c) and (e)-(f)). The full spectra of rubella resemble that of Figure 3 with a sharp peak at 1 year and a broad multi-annual peak.

For future discussion of the implications of vaccination and decline or increase in birth rates, it is useful to investigate how the spectra of rubella change with $R_0$. Keeping the amplitude of seasonal forcing and infectious period of rubella fixed and increasing $R_0$, we expect the period of stochastic fluctuations as well as their amplification to decrease. This is seen from Figure 5 and also illustrated in Figure 7 (a) where the full spectra for parameters close to rubella estimates are shown. The relative contribution of multi-annual and annual frequency components in model time series can be read from the same figure. For small $R_0$ the fluctuations are large and the multi-annual peak is dominant but for larger $R_0$ it becomes subdominant and the annual peak is enhanced. The increase of $\epsilon$ (as well as the increase of population size as discussed before in the text accompanying Figure 4) results in the enhancement of the deterministic peak too (compare Figure 5 (g) and (h)), but does not change the dominant period of fluctuations significantly.

For measles, there are major changes in the behavior as both coherence and amplification increase drastically for $\epsilon>0$, see the newly appeared oval-shaped regions near measles parameters in Figure 5 (b)-(c) and (e)-(f). To demonstrate that this phenomenon indicates the appearance of a biennial cycle in simulations we show in Figure 7 (b) the full spectra for fixed infectious period, $1/\nu=16$ d, and a range of $R_0 \in [22,29]$. These values of $1/\nu$ and $R_0$ are slightly higher than the commonly used estimates for measles in England and Wales, e.g. $1/\nu=13$ d and $R_0=18$ for London before vaccination. The same qualitative dynamics is observed for lower values if the simulation length is shorter than 500 years (results not shown). For $R_0=22$ the spectra are typical of fluctuations about an annual limit cycle with a deterministic peak at 1 year and a broad stochastic peak near 2 years. If $R_0$ is increased further the fluctuations around an annual cycle become macroscopic (the stochastic peak at 2 years becomes much higher) smoothly turning into a biennial limit cycle. This transition corresponds to a period doubling bifurcation in the deterministic model. A strong biennial behavior with a dominant peak at 2 years and a secondary harmonic at 1 year is observed for example, for $R_0=26$. Finally, for even larger $R_0$ we see a transition from biennial to an annual cycle again. The set of transitions seen in Figure 7 (b) is typical of measles and have been observed in related models of infections dynamics via analytical and numerical studies in other research \citep{schwartz:84, schwartz:85, earn:00, blackb:10, rozhnova:10, stone:12}.

The seasonality would act to change the picture in Figure 7 (b) in the following way. From the comparison of Figure 5 (g) and (h) the region of parameter space where such a behavior is seen is expected to get larger with increasing $\epsilon$, in particular for $\epsilon>0.1$ the period doubling transition is induced for values of $R_0$ much smaller than in Figure 7 (b). 

Previous analysis of measles data from England and Wales and the USA has shown that transitions in the dynamics due to an increase or decline of birth rates as well as the introduction of vaccination are associated with transition between annual and biennial limit cycles \citep{earn:00}. Using a simple mapping from changes in vaccination and birth rates to effective changes in $R_0$ introduced in \citep{earn:00}, our results are consistent with this view. For large communities with very high birth rates (high $R_0$) such as Liverpool before vaccination, US cities in the period after the Great Depression or cities in developing countries, we would expect to be in the regime with an annual cycle \citep{finkenstadt:98, finkenstadt:00, earn:00, conlanBirth:2007}. Other large cities with smaller birth rates such as London are in the regime with a biennial cycle \citep{keelinggrenfell:02}. The corresponding spectrum with narrow and sharp peaks at 2 and 1 years has been the main reason of more regular and thus more predictable patterns of measles epidemics in large cities. The vaccination introduced in UK in 1968 lowered $R_0$ and induced a transition from the biennial to the annual cycle with large stochastic fluctuations. Our analysis thus offers an insight into the factors responsible for the shift from regular epidemics of measles before vaccination to less irregular in the vaccine era \citep{rohani:99}.

The last finding deserving a further comment concerns the switching between an annual and triennial cycles found for moderate values of $R_0$, see Figure 5 (e) and Figure 6. This behavior may be responsible for the triennial cycles observed in Baltimore and other US cities during the Great Depression \citep{london:73} but more thorough analysis is needed to confirm this.

\begin{figure*}
\centering
\includegraphics[trim=0cm 0cm 0cm 0cm, clip=true, width=0.48\textwidth]{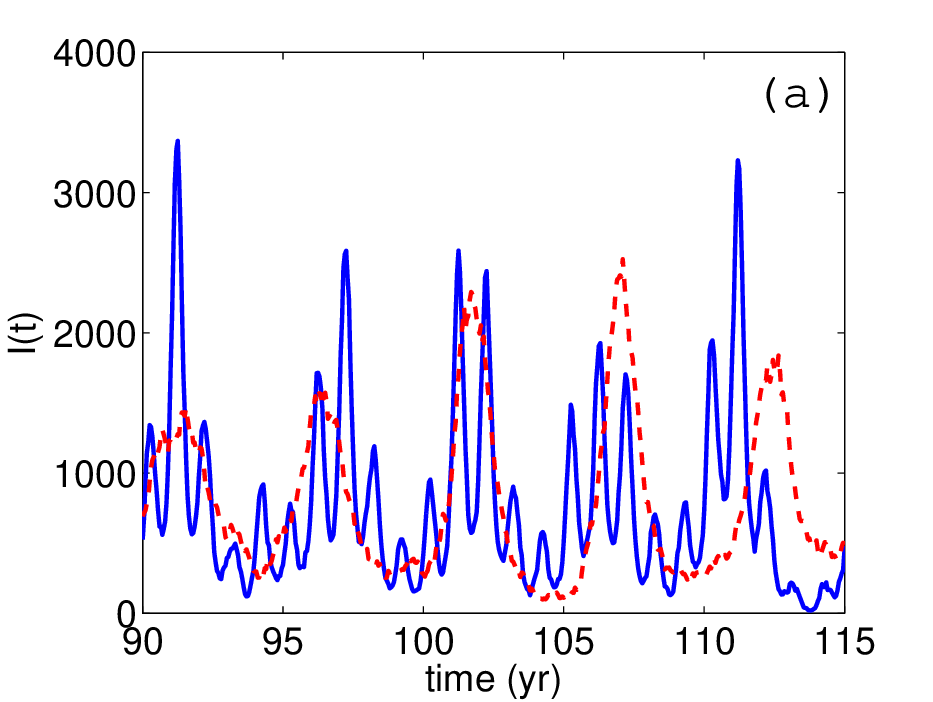}
\includegraphics[trim=0cm 0cm 0cm 0cm, clip=true, width=0.48\textwidth]{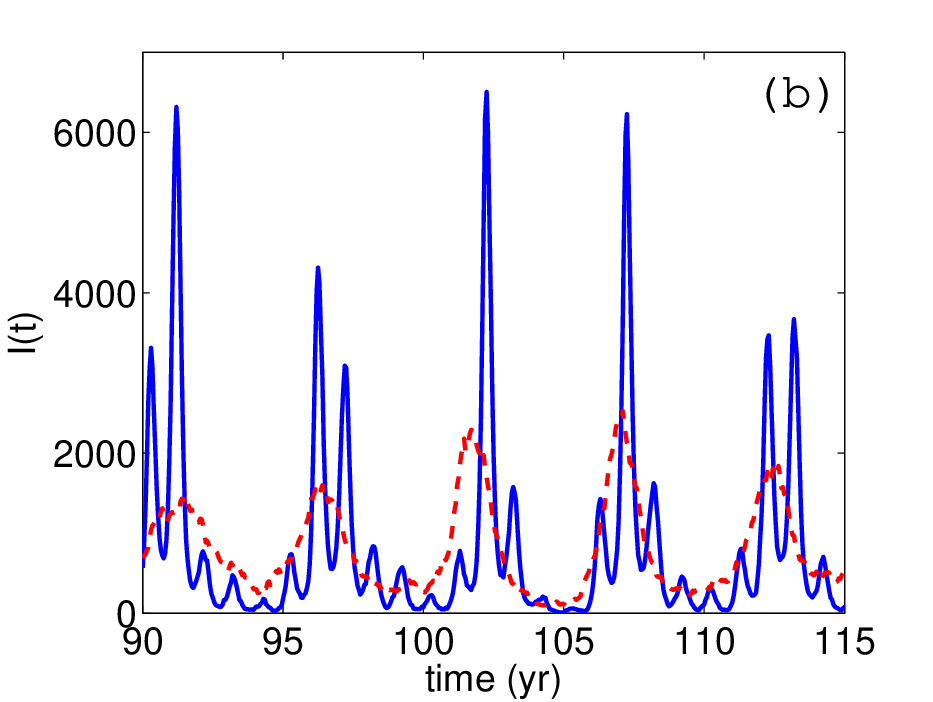}
{\caption{(Color online) Typical time series for rubella parameters. Parameters: $N=10^6$, $\mu=0.02$ 1/y, $R_0=5$, $1/\nu=18$ d, $\epsilon=0$ (dashed red line), (a) $\epsilon=0.2$ (solid blue line), (b) $\epsilon=0.3$ (solid blue line).}}
\label{fig8}
\end{figure*}

\section{Discussion}

In this paper, we have investigated the behavior of the stochastic seasonally forced SIR model based on spectra of long time series for a large range of basic reproductive ratios and infectious periods. For relatively low values of $R_0$ relevant for rubella the model predicts spectra with a stochastic multi-annual peak at about 5-6 years and a deterministic annual peak. Both peaks are observed in the spectra of rubella data (Figure 1). The multi-annual peak stays largely unchanged under the introduction of seasonality (Figures 3 and 5 (a), (d)) or population size (Figure 4) which explains its presence in time series from different locations. 

Using the complementary measures, coherence and amplification, we further studied how the structure of stochastic fluctuations in the model changes with $R_0$. By definition of these measures (Section 2) it is not possible to estimate them from a single (data) time series such as shown in Figure 1. We therefore cannot compare our model's predictions for these measures to the data directly. However, it is still possible to match the data and model's full spectra (i.e. not only the positions of the stochastic and deterministic peaks but also their heights) given the information about the rate of reporting and the population size. We did this for the Mexico time series (see Figure 4 in the Supplementary Material) and obtained a very good agreement between the model and the data. For the Canada and Japan time series we lack the aforementioned information so these analyses yield the full spectra with correct positions of the peaks (as discussed and shown in Section 3.2.2) but their heights can vary depending on the reporting rate and population size used in the model.   

\begin{figure*}
\centering
\includegraphics[trim=0cm 0cm 0.6cm 0.4cm, clip=true, width=0.48\textwidth]{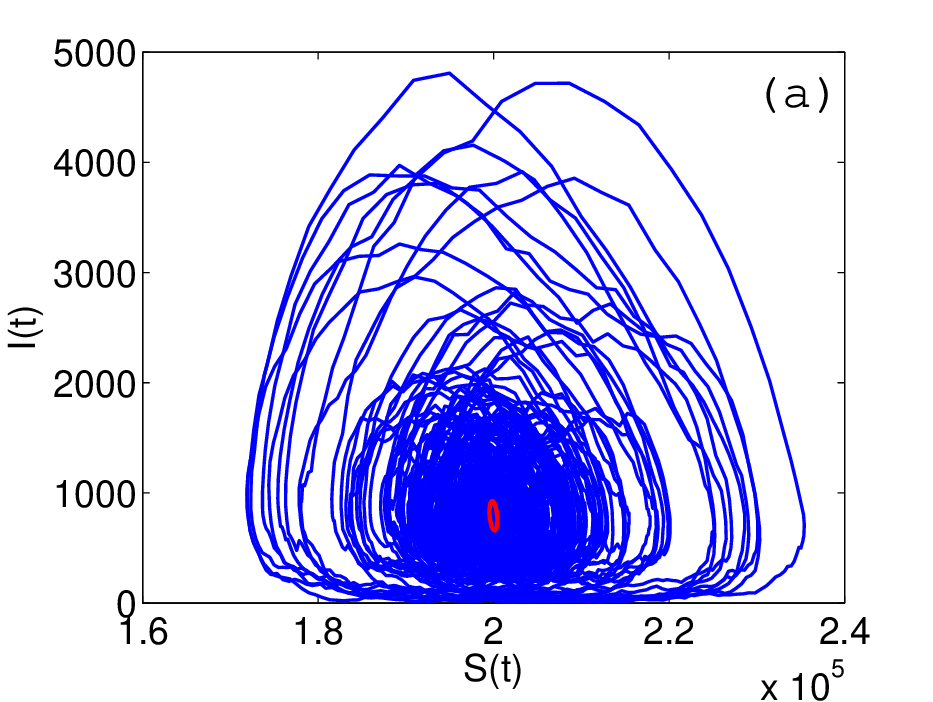}
\includegraphics[trim=0cm 0cm 0.6cm 0.4cm, clip=true, width=0.48\textwidth]{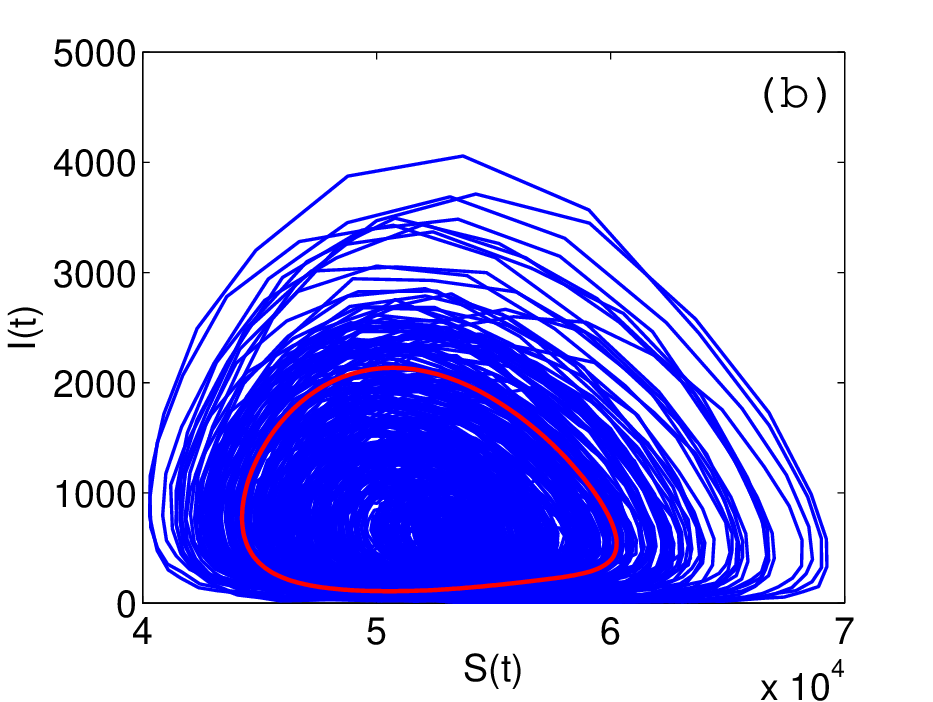}
{\caption{(Color online) Typical stochastic trajectories from simulation (blue solid lines) shown in the susceptible-infected plane. The red line is the deterministic annual cycle in (a) and the biennial cycle in (b). For the parameters used in this figure many simulations go extinct quite fast, for illustration purposes we have chosen the ones which lasted for 500 years. Parameters: $N=10^6$, $\epsilon=0.05$, $\mu=0.02$ 1/y, (a) $R_0=5$, $1/\nu=18$ d (close to rubella estimates), (b) $R_0=19$, $1/\nu=12$ d (close to measles estimates).}}
\label{fig9}
\end{figure*}

A visual inspection of simulated time series demonstrates intriguing behavior emerging from the interaction between stochasticity and a deterministic annual cycle. Figure 8 shows typical time series for the unforced (red dashed line) and seasonally forced (blue solid line) cases. If $\epsilon=0$ the epidemic patterns represent multi-annual coherent oscillations. As $\epsilon$ is increased we find qualitatively different dynamics all of which correspond to spectra with a multi-annual and an annual peak. Figure 8 (a) shows an example of annual epidemics of alternating amplitudes modulated by an oscillation of a long period corresponding to the period of stochastic multi-annual fluctuations. This dynamic is qualitatively similar to the multi-annual regularity observed, for example, for rubella in Japan (Figure 1). We also find very large outbreaks followed by outbreaks of much lower amplitude as in Figure 8 (b). Such a behavior may be responsible for the spiky dynamics observed in, for example, Canada (Figure 1). Note that the spikes in the data could also arise from spatial effects such as local extinction of the disease followed by reintroduction from another region. However, as we do not possess more resolved data it is impossible to reach a final conclusion with regards to this issue.

The patterns of rubella incidence in large populations are in contrast with those of measles. For the latter the spectra are characterised by sharp and narrow peaks at 1 and 2 years (as opposed to the broad multi-annual peak and a narrow peak at 1 year observed for rubella) and thus correspond to much more regular dynamics. The transitions in measles behavior due to vaccination or change in birth rates are associated with transitions between the annual and biennial limit cycles of the deterministic model. In future work, it would be interesting to study stochastic measles dynamics for higher levels of seasonal forcing that correspond to chaos in the deterministic model \citep{ferrari:08}.

Both measles and rubella are found to be close to the extinction boundary, and increasing the amplitude of seasonal forcing only extends the region of parameter space with frequent extinctions. Figure 9 illustrates typical stochastic trajectories from simulation in the susceptible-infected plane from which the spectra were computed. Interestingly, these patterns suggest that the mechanism accounting for high extinction probability is different for rubella and measles. For rubella, extinction occurs as a consequence of large stochastic fluctuations about a small (and globally less stable) annual limit cycle, see Figure 9 (a). For measles, extinctions are mainly due to the shape of a large (and globally more stable) biennial limit cycle, from which the system can be driven to extinction by even relatively small fluctuations, see Figure 9 (b). 

We can use the framework developed to predict the effect on persistence of an effective reduction in $R_0$ by vaccination or declining birth rates for rubella and measles. For measles in the biennial regime, either an increase or a decrease of $R_0$ can lead to fluctuations around an annual cycle (rather than a biennial cycle) that could result in lower extinction rates and thus higher persistence. For rubella, a reduction in $R_0$ will lead to larger and more coherent oscillations that would unambiguously result in higher extinction probabilities, and thus lower persistence. Both these outcomes merit serious consideration in a public health context: vaccination against measles can make local elimination less likely \citep{ferrari:08}; while vaccination against rubella is likely to increase local extinction, allowing the buildup of susceptible individuals in later age classes \citep{metcalf:2011,metcalf:2011a}, potentially leading to an increase in the burden of Congenital Rubella Syndrome.

These conclusions point to a need for theoretical developments towards uncovering the mechanisms of stochastic extinctions in small population based on the analysis of epidemic models (a thorough overview of studies in this area aiming at understanding the persistence of measles is given in the recent work by Conlan et. al. \citep{conlan:10}). In the mathematical framework we adopted in this paper, the approach to computation of the distribution of extinction times in an unforced stochastic epidemic model was proposed some time ago \citep{nasell:99, nasell:05}. The disease persistence in stochastic epidemic models can also be studied using the so-called WKB approximation but the method is only applicable to low-dimensional and unforced models (e. g. \citep{Black:wkb,schwartz:2011}). Nevertheless, no analytical progress can be made along the same lines for the seasonally forced model we use here. We are aware of only one study \citep{bacaer} which addressed extinction probabilities in the periodic context using theoretical methods but the method of \citep{bacaer} has a limitation because it applies to the large population limit only. The development of the approaches to compute the time to extinction in seasonally-forced models will be therefore a subject of further research.

Our focus has been on measles and rubella; however, the broad span of parameter space explored means that our results may shed light on the dynamics of other diseases whose dynamics can be described by a simple SIR formalism with seasonal forcing; for example pertussis \citep{nguyen:08,rozhnova:12}. Although infection with pertussis does not confer permanent immunity the SIR model has been shown to capture the qualitative patterns to some extent \citep{rozhnova:12}. Taking the pertussis parameters before vaccination that are well established in independent data sources (see e.g. \citep{anderson:1991,bauch:2003}, $1/\nu=22$ d and $R_0$=17 for London and Ontario, Canada) from Figure 5 we see that for pertussis the dominant period of stochastic fluctuations is 2 to 3 years. These periods are in the agreement with the documented interepidemic periods \citep{rozhnova:12,bauch:2003}. We also find that the coherence of fluctuations is very low, which is compatible with the famously irregular dynamics of pertussis. The decrease of $R_0$ due to vaccination would act to increase the dominant period and coherence which also agrees with the observation of the shift to more regular dynamics in the vaccine era \citep{rohani:99}.

To conclude, in this paper, we used a stochastic framework to explain the recurrent dynamics of rubella, particularly in comparison to measles. Our analysis revealed that whilst both rubella and measles are relatively close to their extinction boundary, the reasons for this are very different. Finally, our analysis showed that, for rubella, reducing $R_0$ by vaccinating or a declining birth rate unambiguously result in higher extinction probabilities, whereas, for measles, outcomes can be more complicated; and both these facts have public health implications.\newline

{\bf Acknowledgements}\newline

We thank Kihei Terada for helpful discussion of rubella dynamics in Japan; Alan McKane for his careful reading and comments on the paper; and Ottar Bj{\o}rnstad for the help with the computation of the confidence intervals for data spectra. Rozhnova gratefully acknowledges the financial support from the Portuguese Foundation for Science and Technology under Contract No. SFRH/BPD/69137/2010. Grenfell was supported by the RAPIDD program of the Science \& Technology Directorate, Department of Homeland Security and the Fogarty International Center, National Institutes of Health. Metcalf and Grenfell were supported by the Bill and Melinda Gates Foundation.

\bibliographystyle{plain}



\end{document}


\bigskip

\centerline{\Large \textbf{Supplementary Material}}

\vspace{1cm}

{\large{\bf Characterizing the dynamics of rubella relative to measles:}}

\centerline{\large{\bf the role of stochasticity}}

\vspace{0.7cm}

\centerline{\large{\textit{Ganna Rozhnova\footnote{corresponding author: ganna.rozhnova@manchester.ac.uk}, C. Jessica E. Metcalf and Bryan T. Grenfell}}}

\vspace{0.4cm}

\section{Analytical spectrum}

In this section we describe how to derive the analytical power spectrum for the stochastic SIR model. We use the method known in the physics literature \cite{kampen:1981} as the van Kampen or system size (total population size, $N$) expansion. The discussion of this standard method in the context of compartmental epidemiological models such as the SIR model we consider here or other related models of infectious diseases can be found in \cite{alonso:07,blackb:10,rozhnova:10,rozhnova:11,rozhnova:12}. The most recent description of the method in application to the SIRSI model is given in the supplemental information to Ref. \cite{rozhnova:12}. Refs \cite{alonso:07,blackb:10} outline the derivation for the SIR model with immigration and term-time forcing. Here we briefly revise the steps of the calculation for this model with sinusoidal forcing and without imports, see Sec. 2.1 of the main text.

We begin by listing the transition rates, $\mathcal{T}(\sigma'|\sigma)$, between an initial state $\sigma$ and a final state $\sigma'$ of the system, where $\sigma=\{S(t),I(t)\}$. $S(t)$ and $I(t)$ are the numbers of susceptible and infectious individuals at time $t$. We consider the case of linked birth and death events, so that $N$ is constant and the number of recovered individuals is given by $R(t)=N-S(t)-I(t)$. The transition rates associated to the processes of infection, recovery, birth and death read as follows:

\begin{itemize}
	\item Infection: $S\stackrel{\lambda}{\longrightarrow} I$, where $\lambda=\beta(t)I(t)/N$ is the force of infection and $\beta(t)=\beta_0(1+\epsilon\cos 2\pi t)$ is a sinusoidally varying transmission rate:
\begin{equation}
\label{infection}
\mathcal{T}(S-1,I+1|S,I)=\beta(t)I(t)/N S(t).
\end{equation}
	\item Recovery: $I\stackrel{\nu}{\rightarrow} R$, where $\nu$ is the rate of recovery:
\begin{equation}
\label{recovery}
\mathcal{T}(S,I-1|S,I)=\nu I(t).
\end{equation}
	\item Birth and death: $R\stackrel{\mu}{\rightarrow} S$ and $I\stackrel{\mu}{\rightarrow} S$, where $\mu$ is the rate of birth and death:
\begin{equation}
\label{birthdeath1}
\mathcal{T}(S+1,I|S,I)=\mu [N-S(t)-I(t)],
\end{equation}
\begin{equation}
\label{birthdeath2}
\mathcal{T}(S+1,I-1|S,I)=\mu I(t). 
\end{equation}
\end{itemize}

Note that in some Refs. the birth and death events are not linked (e.g. \cite{simoes:08, rozhnova:10,rozhnova:12}), that is why the transition rates for these processes are different from Eqs. (\ref{birthdeath1})-(\ref{birthdeath2}) and so the final expressions for the spectrum too. Stochastic simulations of the system, defined by Eqs. (\ref{infection})-(\ref{birthdeath2}), based on the time-dependent extension \cite{anderson:07} of Gillespie's algorithm \cite{gillespie:76}, produce exact realisations of the stochastic process. From a sample of such simulations we compute numerical spectra and their characteristics (amplification, coherence and the main period), as described in the main text.

The underlying stochastic process is Markovian because the transition rates, Eqs. (\ref{infection})-(\ref{birthdeath2}), depend only on the present state and not on the previous states. The full dynamics of this system is completely described by the master equation for the probability distribution of having the system in state $\sigma$ at time $t$ \cite{kampen:1981}:

\begin{equation}
\label{masterequation}
\frac{d\mathcal{P}(\sigma,t)}{dt}=\sum\limits_{\sigma'\neq \sigma} \mathcal{T}(\sigma|\sigma')\mathcal{P}(\sigma',t)-\sum\limits_{\sigma'\neq \sigma}\mathcal{T}(\sigma'|\sigma)\mathcal{P}(\sigma,t).
\end{equation}
\\
\noindent{}The approximate analytical spectrum of the fluctuations, ${\bf x}=\{x_S(t),x_I(t)\}$, around the densities of susceptible, $\bar{s}$, and infective, $\bar{i}$, individuals, can be derived from the master equation for large but finite $N$ \cite{kampen:1981} by making the substitutions 

\begin{equation}
S(t)=N\bar{s}+\sqrt{N}x_S(t), \ \ \ I(t)=N\bar{i}+\sqrt{N}x_I(t)
\end{equation}
\\
\noindent{}into Eqs. (\ref{infection})-(\ref{birthdeath2}), inserting these into Eq. (\ref{masterequation}) and expanding it in powers of $1/\sqrt{N}$. This leads to the equation in which terms multiplied by different powers of $1/\sqrt{N}$ appear. The leading order terms are proportional to $\sqrt{N}$, then the terms of order $N^0$, etc. The leading order terms give rise to the seasonally-forced deterministic SIR model for the densities of susceptible and infective individuals: 

\begin{eqnarray}
\dfrac{d\bar{s}}{dt}&=&\mu\left(1-\bar{s}\right)-\beta(t) \bar{s}\bar{i}, \nonumber\\
\label{SandIeq}
\dfrac{d\bar{i}}{dt}&=&\beta(t) \bar{s}\bar{i}-(\nu+\mu)\bar{i}.
\end{eqnarray}
\\
At the next to leading order we obtain a linear Fokker-Planck equation for the probability distribution $\Pi({\bf x},t)$ \cite{kampen:1981}:

\begin{equation}
\label{fokkerplanckequation}
\dfrac{\partial\Pi}{\partial t}=-\sum\limits_{k,j}{A_{kj}(t)\dfrac{\partial (x_j\Pi)}{\partial x_k}}+\frac{1}{2}\sum\limits_{k,j}{B_{kj}(t)\dfrac{{\partial}^2\Pi}{\partial x_k \partial x_j}}, \ \ \ k,j=S,I.
\end{equation}
\\ 
Since the above equation is linear its solution, $\Pi({\bf x},t)$, is a multivariate Gaussian distribution completely determined by the first and the second moments \cite{kampen:1981}. The matrices ${\bf A}(t)$ and ${\bf B}(t)$ in this equation are time-dependent through $\bar{s}$, $\bar{i}$ and $\beta(t)$ in the case of seasonal forcing ($\epsilon>0$), and time-independent when the forcing is absent ($\epsilon=0$). Carrying out the calculation shows that ${\bf A}(t)$ is the Jacobian of Eq. (\ref{SandIeq}) and ${\bf B}(t)$ is the symmetric matrix which must be obtained directly from the expansion. When $\epsilon=0$, we evaluate the Jacobian and matrix ${\bf B}(t)$ at the non-trivial fixed point of Eq. (\ref{SandIeq}) (with $\epsilon=0$), and Eq. (\ref{fokkerplanckequation}) describes the fluctuations about the endemic steady state. For $\epsilon>0$, the matrices ${\bf A}(t)$ and ${\bf B}(t)$ are evaluated on the attractors of Eq. (\ref{SandIeq}), which are limit cycles with periods multiples of 1 year, and are thus periodic with the period of the cycle. In this case Eq. (\ref{fokkerplanckequation}) describes the fluctuations about the cycle, around which the expansion is performed.

The power spectrum of the fluctuations for infectives is most readily obtained from the Langevin equation \cite{kampen:1981} to which Eq. (\ref{fokkerplanckequation}) is equivalent:

\begin{equation}
\label{langevinequation}
\dfrac{d{\bf x}(t)}{dt}={\bf A } (t) {\bf x}(t)+ {\bf f}(t),
\end{equation}
\\
\noindent{}where ${\bf f}(t)$ are Gaussian white noise terms with correlator $\left\langle {\bf f}(t){\bf f}(t')\right\rangle={\bf B}(t)\delta(t-t')$. 

For $\epsilon=0$, the spectrum about the non-trivial fixed point is defined as the averaged squared modulus of the Fourier transform of $x_I(t)$ (see e.g. \cite{alonso:07,rozhnova:12,rozhnova:11}). Eq. (\ref{langevinequation}) is linear so that the calculation of the Fourier transform is straightforward. We give here the final expression for the spectrum \cite{rozhnova:12,simoes:08}:

\begin{equation}
P_I(\omega)=\dfrac{2\mu (\beta_0-\nu-\mu)}{Z\beta_0}\left(\omega^2+\dfrac{(\mu\beta_0)^2}{(\nu+\mu)^2}\right),
\end{equation}
\\
\noindent{}where $Z=(D-\omega^2)^2+T^2\omega^2$, $D=\mu(\beta_0-\nu-\mu)$ and $T=-\beta_0\mu/(\nu+\mu)$. The results are shown in the units of $f=\omega/2\pi$, where $\omega$ is the cyclic frequency.

For $\epsilon>0$, the calculations are more evolved and have to be done numerically because no closed expression for a limit cycle can be found and because of the time-dependence of ${\bf A}(t)$ and ${\bf B}(t)$. Recurring to the Floquet theory, the spectrum can be computed as the Fourier transform of the autocorrelation function of the fluctuations $x_I(t)$. We refer the reader to \cite{blackb:10,rozhnova:10} for the mathematical details of this theory necessary to perform the calculation. If several stable coexisting limit cycles are present for a given set of parameters, the spectrum of fluctuations around each of them can be obtained separately.
\bibliography{vaccination}
\bibliographystyle{plain}
\newpage
\section{Additional results}

\begin{figure}[h]
\centering
\includegraphics[trim=0.6cm 0.1cm 1.5cm 0.4cm, clip=true, width=0.32\textwidth]{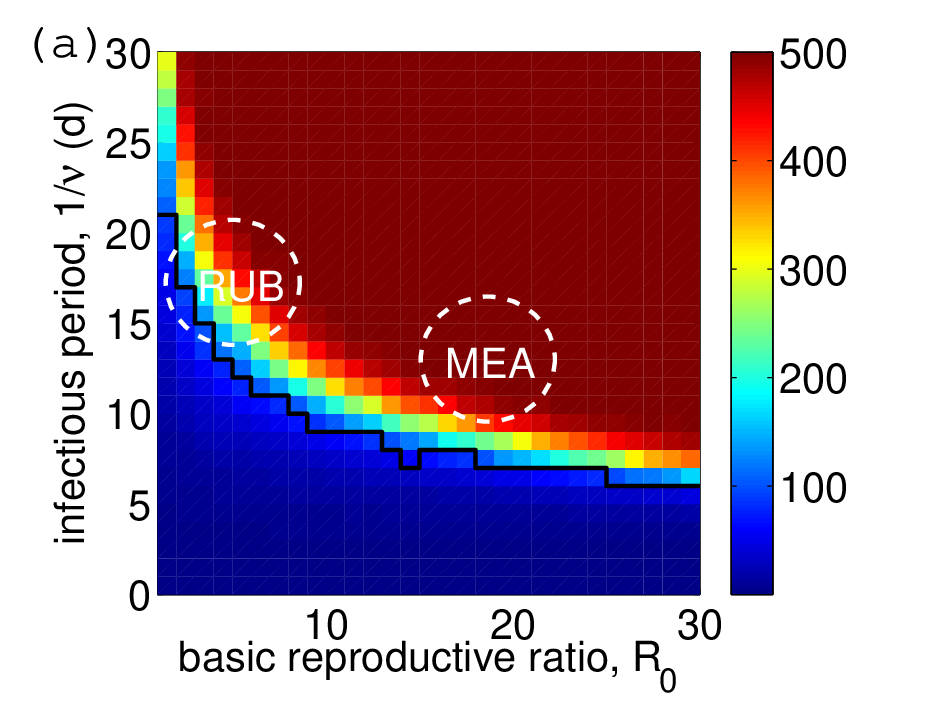}
\includegraphics[trim=0.6cm 0.1cm 1.5cm 0.4cm, clip=true, width=0.32\textwidth]{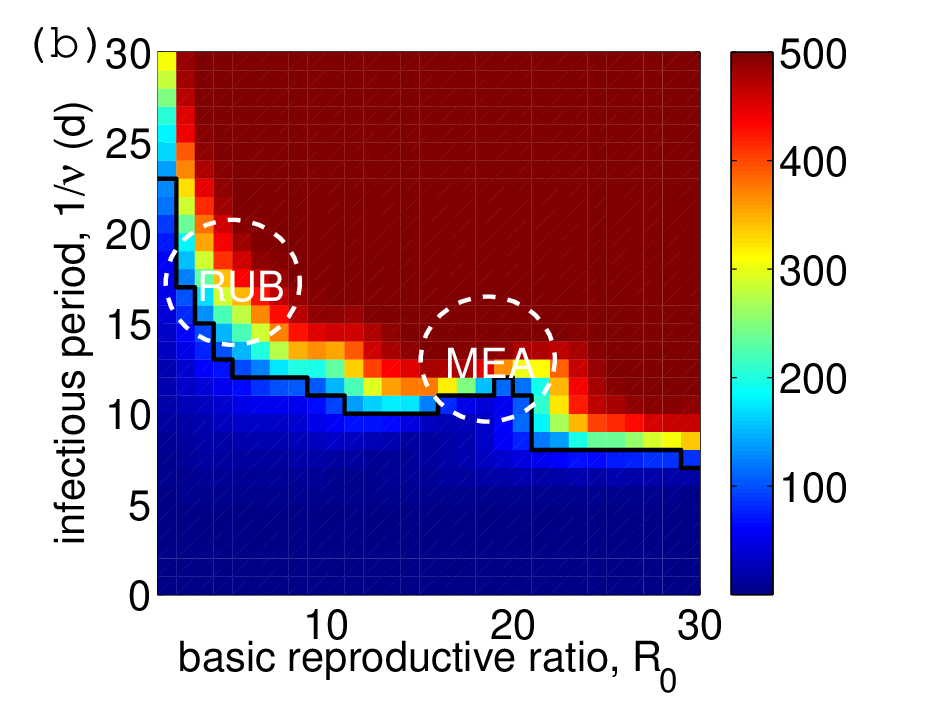}
\includegraphics[trim=0.6cm 0.1cm 1.5cm 0.4cm, clip=true, width=0.32\textwidth]{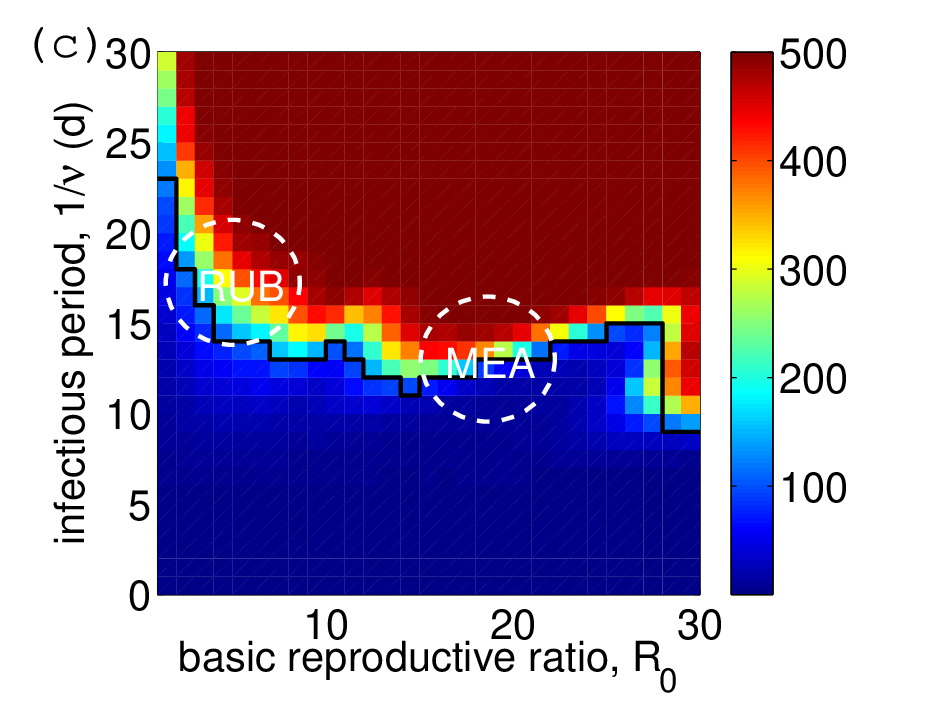}
{\caption{(Color online) The average length of simulations in years for 250 runs with the fixed length of 500 years. Below the black stairstep graph (the extinction boundary) all simulations went extinct within 500 years. In this region the time to extinction is below 200 years. In most of the region above the black line the average length of simulations is 500 years which means that none of the simulations went extinct and that the summary statistics of power spectra estimated from simulations are carried out on 250 runs. For points close to the extinction boundary the statistics are calculated on a variable number of runs. In practice, the spectra computed from as few as 30 runs are very smooth so that the sample size effects become negligible and the discrepancies between simulations and analytic predictions arise purely because of the breakdown of the theory \cite{blackb:10,rozhnova:10,simoes:08}. This is seen from, for example, Figure 4 (b) in the main text which was computed from 30 runs (for parameters 2 points away from the extinction boundary). The sampling errors might appear only for parameters at the very border with the black line, where running long simulations becomes prohibitively computationally intensive. In that region few points may stand out from the general pattern as, for example, seen from the surface plots for amplification and coherence in Figures 3 and 5 of the main text. The same figures show that the dominant period is not affected by sampling. Parameters: $\mu=0.02$ 1/y, $N=10^6$, (a) $\epsilon=0$, (b) $\epsilon=0.05$ and (c) $\epsilon=0.1$.}}
\label{fig1}
\end{figure}

\begin{figure}[h]
\includegraphics[trim=0cm 0cm 0cm 0cm, clip=true, width=0.32\textwidth]{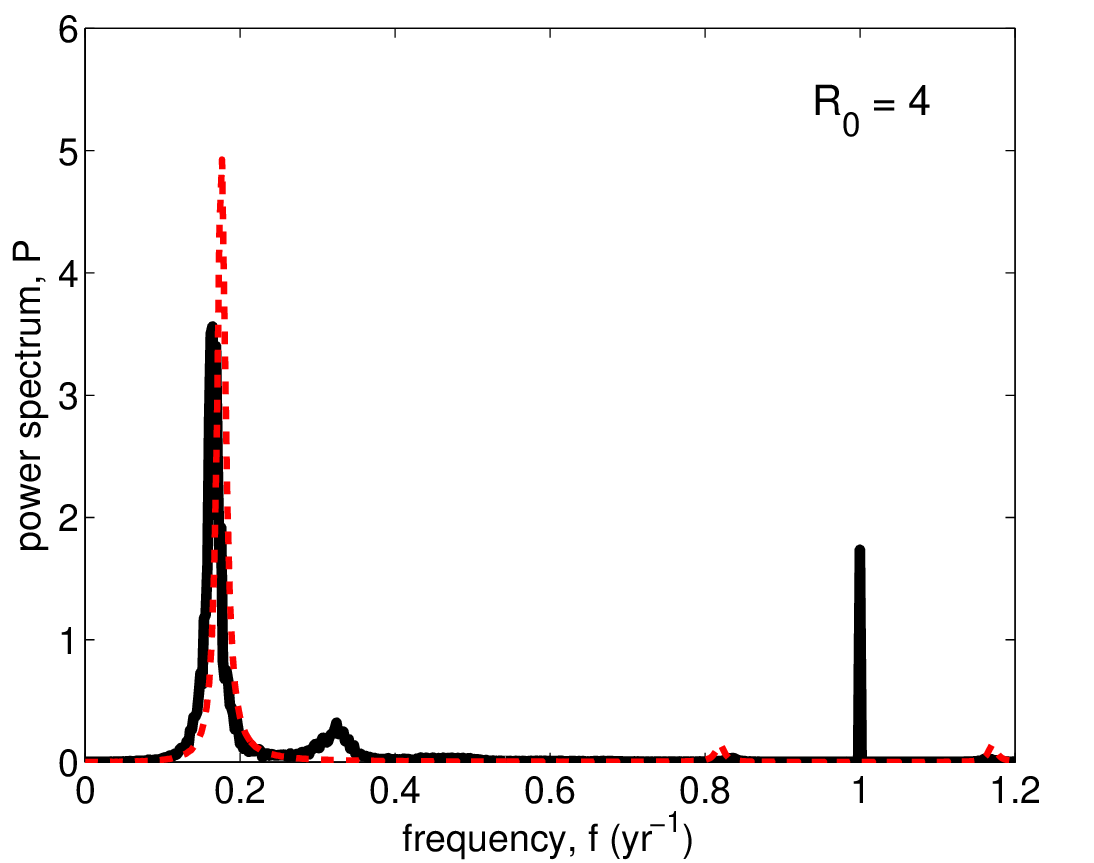}
\includegraphics[trim=0cm 0cm 0cm 0cm, clip=true, width=0.32\textwidth]{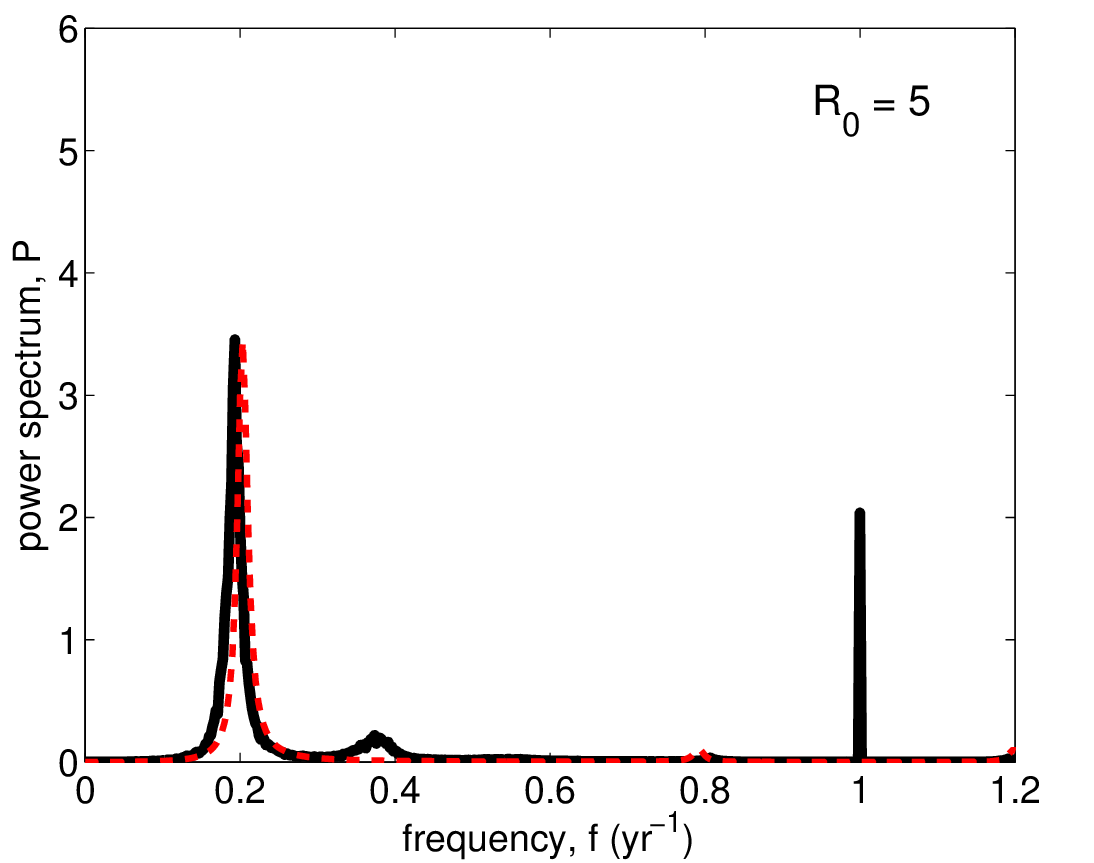}
\includegraphics[trim=0cm 0cm 0cm 0cm, clip=true, width=0.32\textwidth]{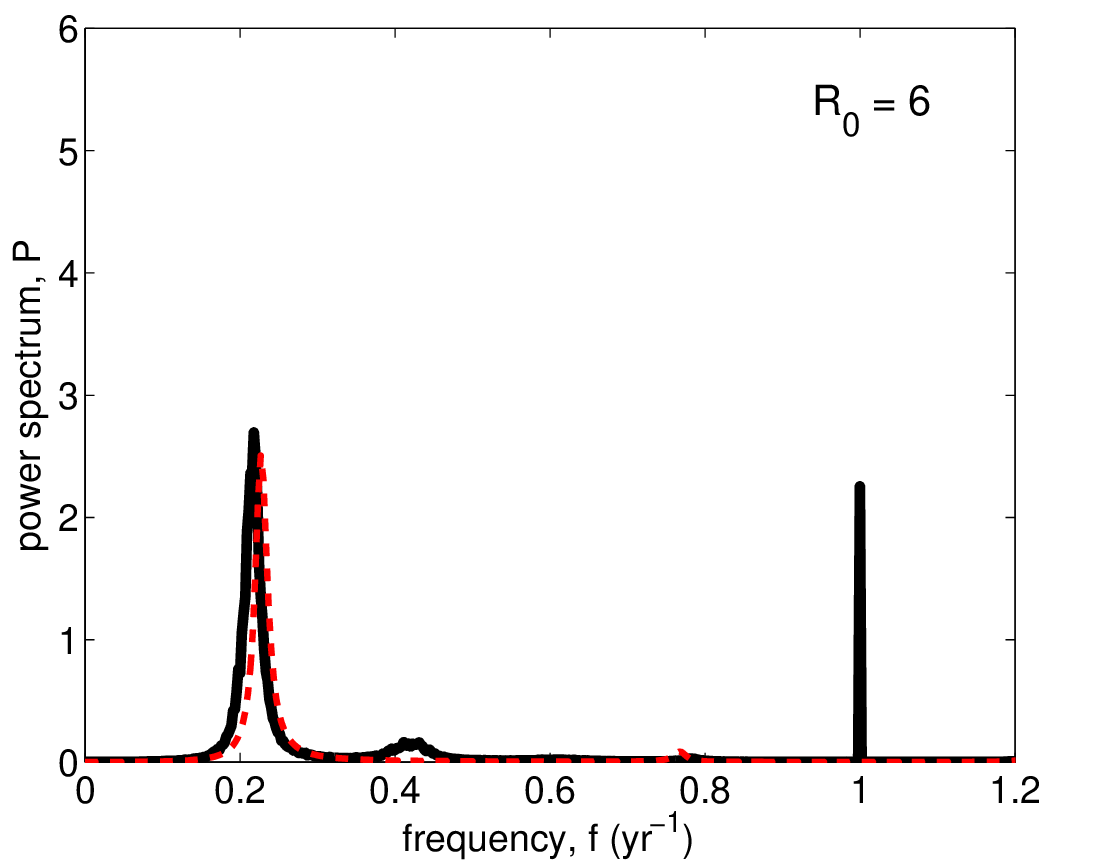}\\
\includegraphics[trim=0cm 0cm 0cm 0cm, clip=true, width=0.32\textwidth]{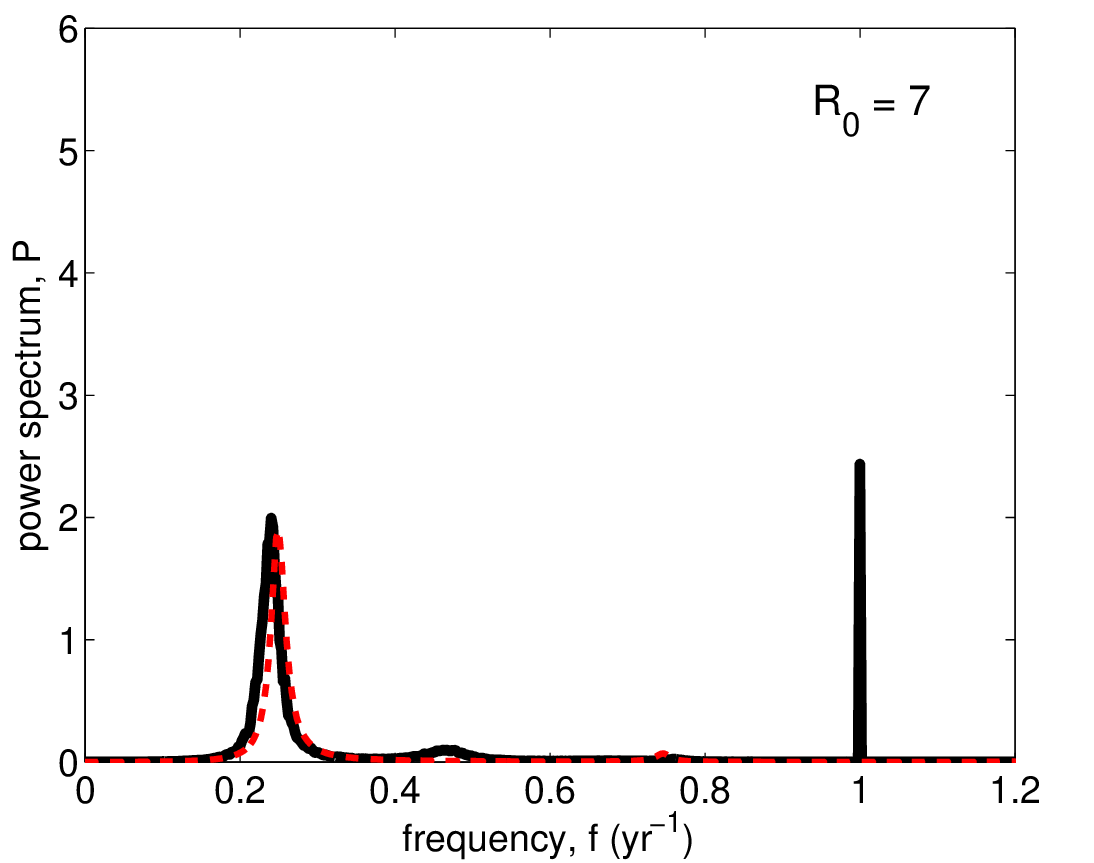}
\includegraphics[trim=0cm 0cm 0cm 0cm, clip=true, width=0.32\textwidth]{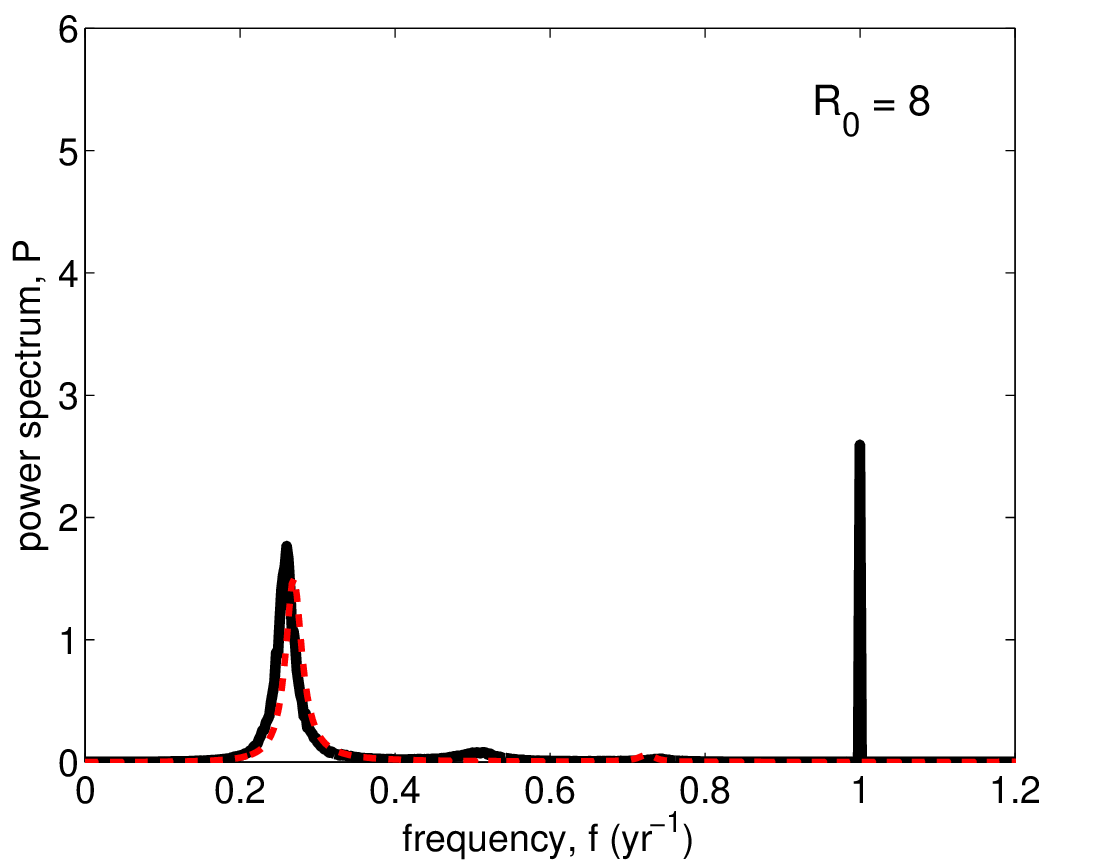}
\includegraphics[trim=0cm 0cm 0cm 0cm, clip=true, width=0.32\textwidth]{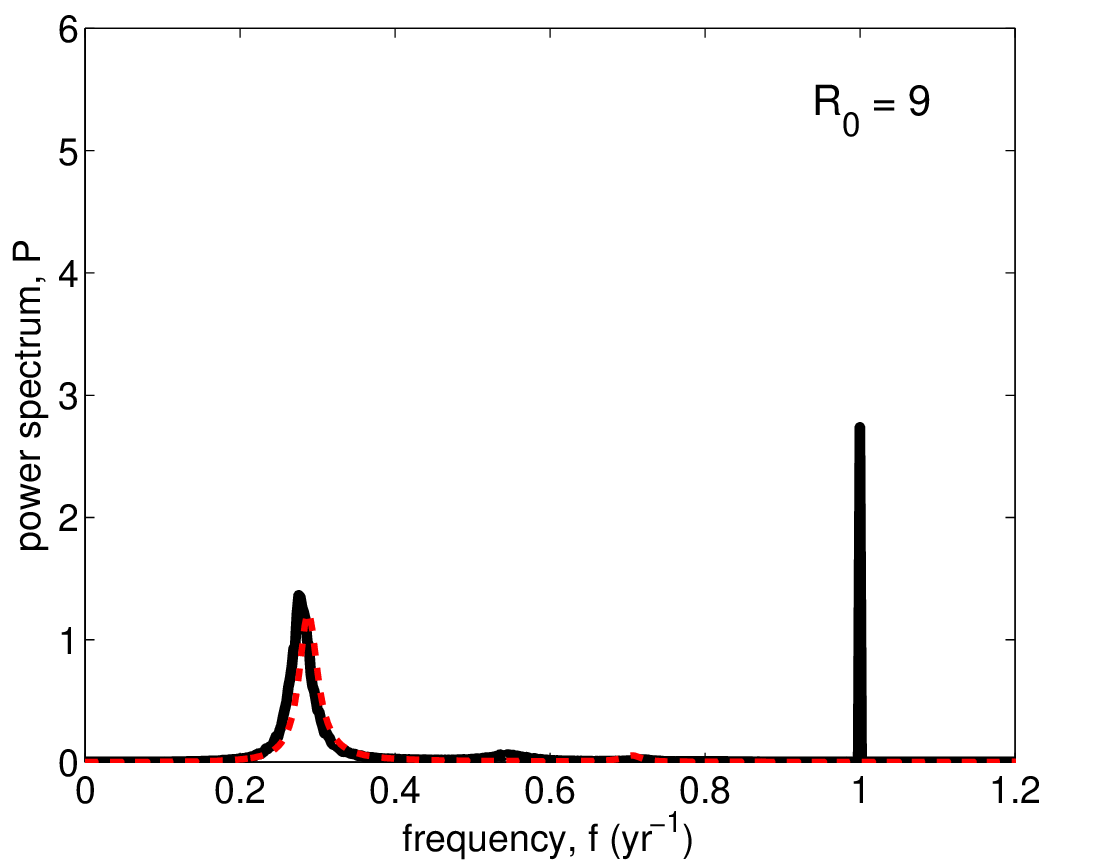}\\
\includegraphics[trim=0cm 0cm 0cm 0cm, clip=true, width=0.32\textwidth]{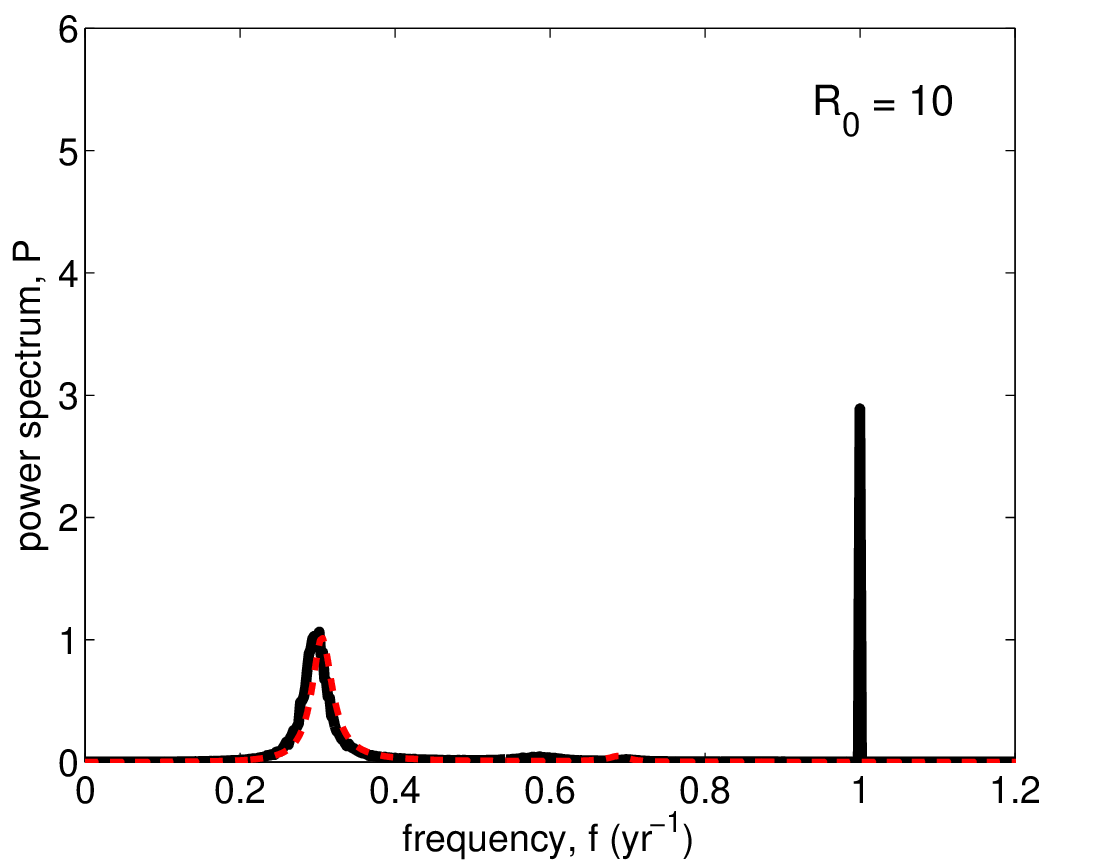}
\includegraphics[trim=0cm 0cm 0cm 0cm, clip=true, width=0.32\textwidth]{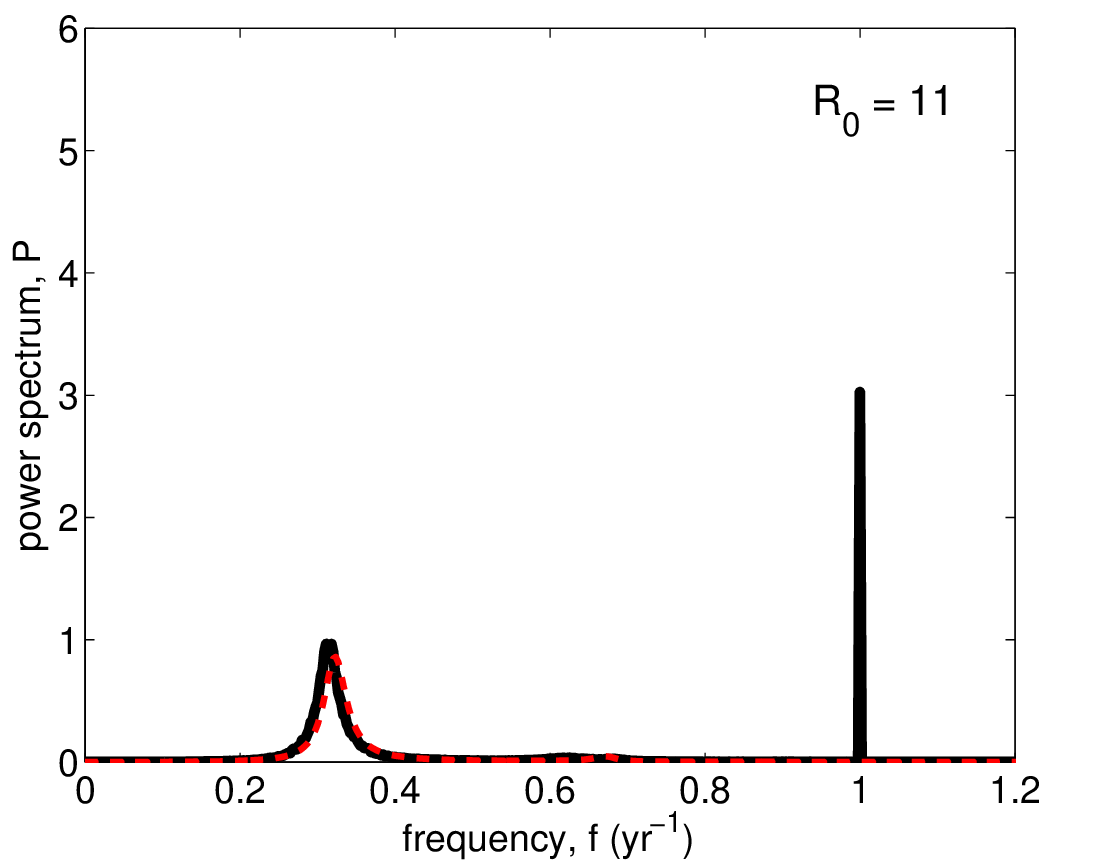}
\includegraphics[trim=0cm 0cm 0cm 0cm, clip=true, width=0.32\textwidth]{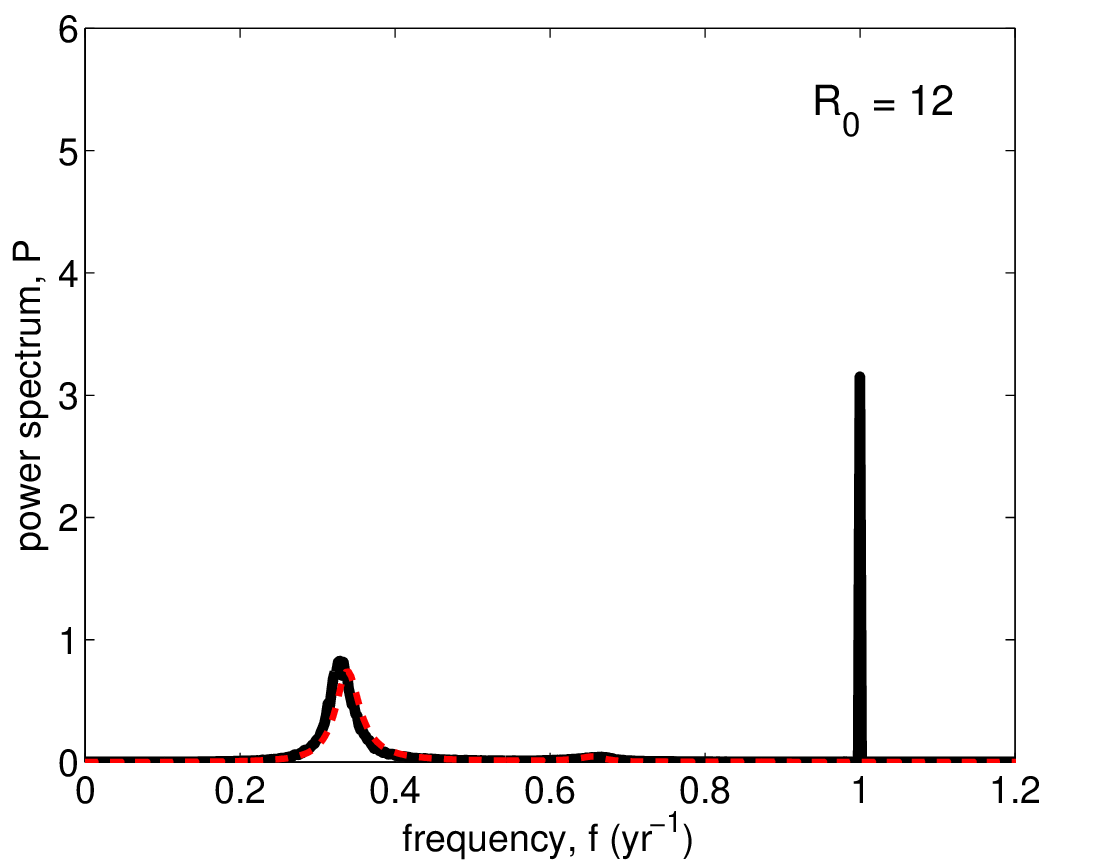}\\
\includegraphics[trim=0cm 0cm 0cm 0cm, clip=true, width=0.32\textwidth]{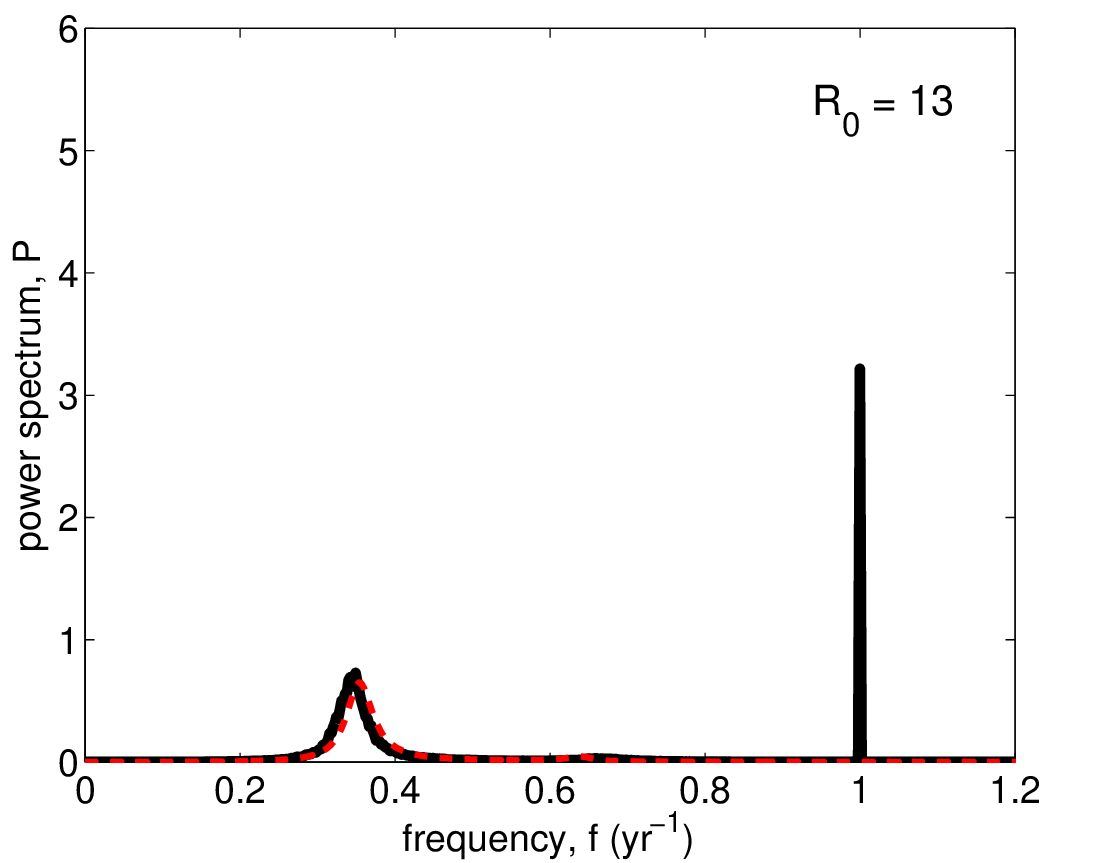}
\includegraphics[trim=0cm 0cm 0cm 0cm, clip=true, width=0.32\textwidth]{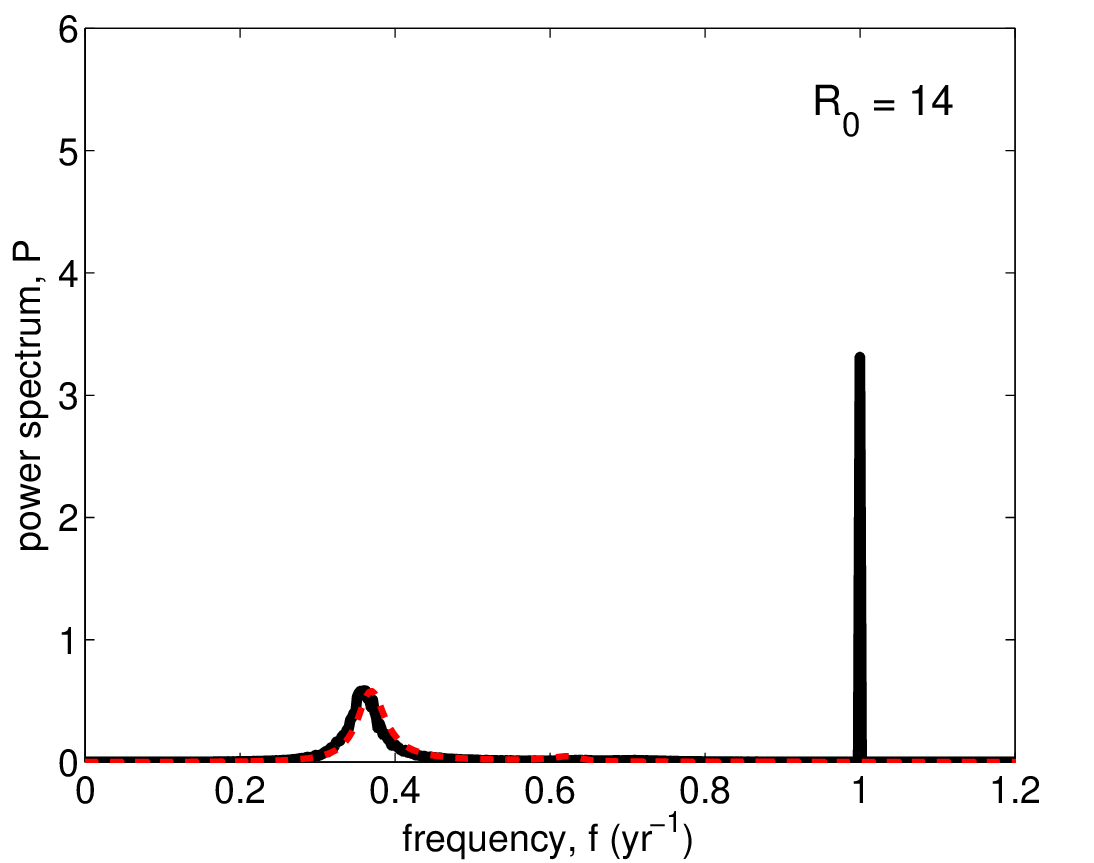}
\includegraphics[trim=0cm 0cm 0cm 0cm, clip=true, width=0.32\textwidth]{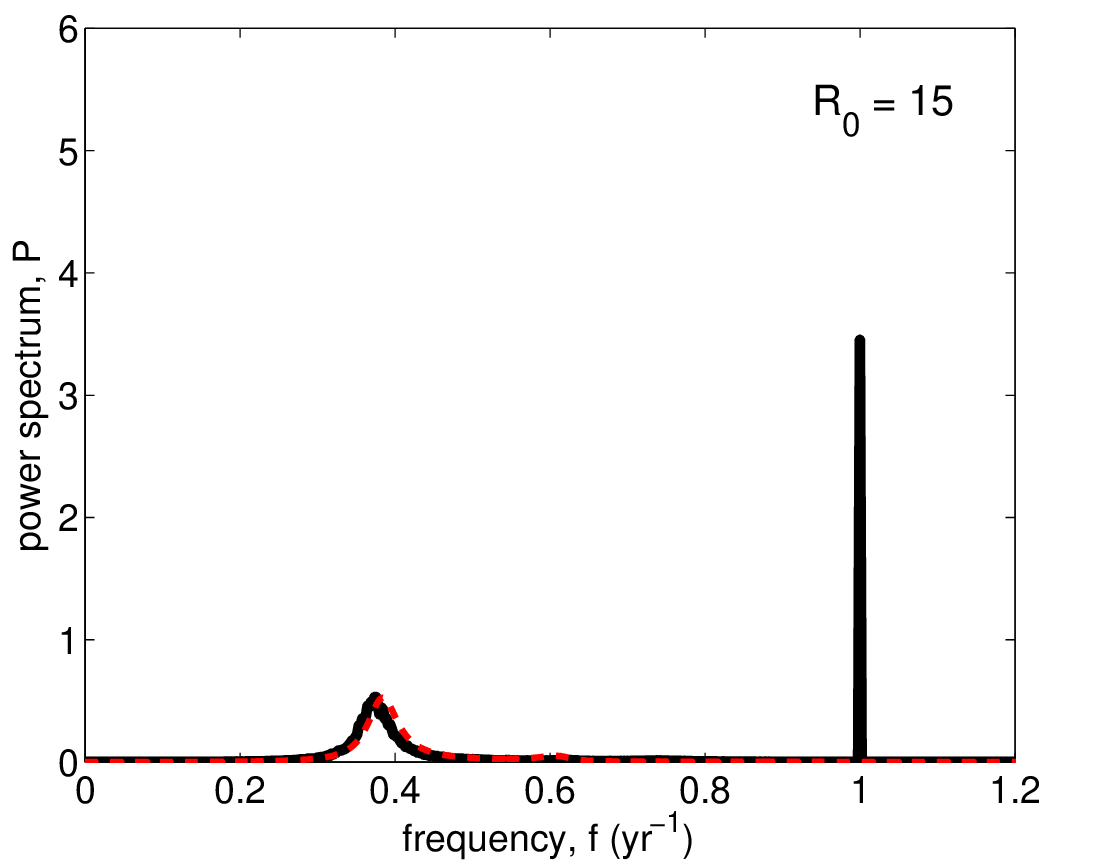}\\
\includegraphics[trim=0cm 0cm 0cm 0cm, clip=true, width=0.32\textwidth]{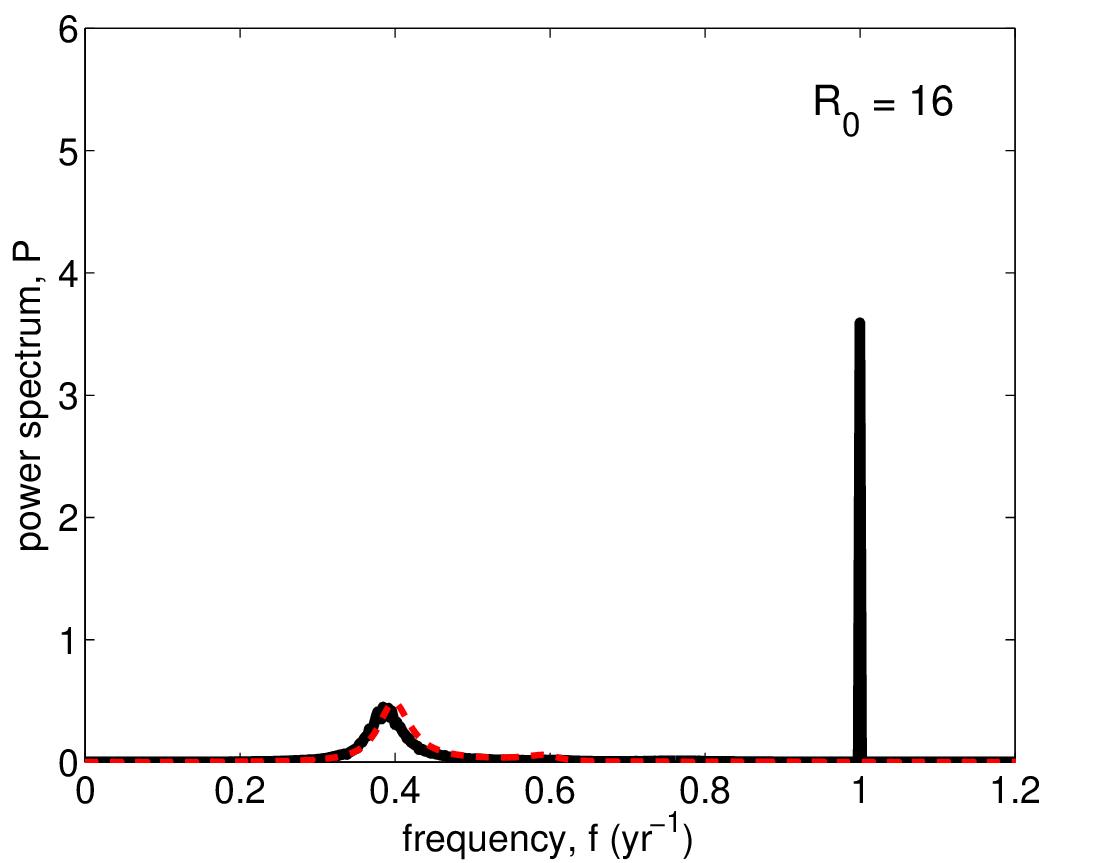}
\includegraphics[trim=0cm 0cm 0cm 0cm, clip=true, width=0.32\textwidth]{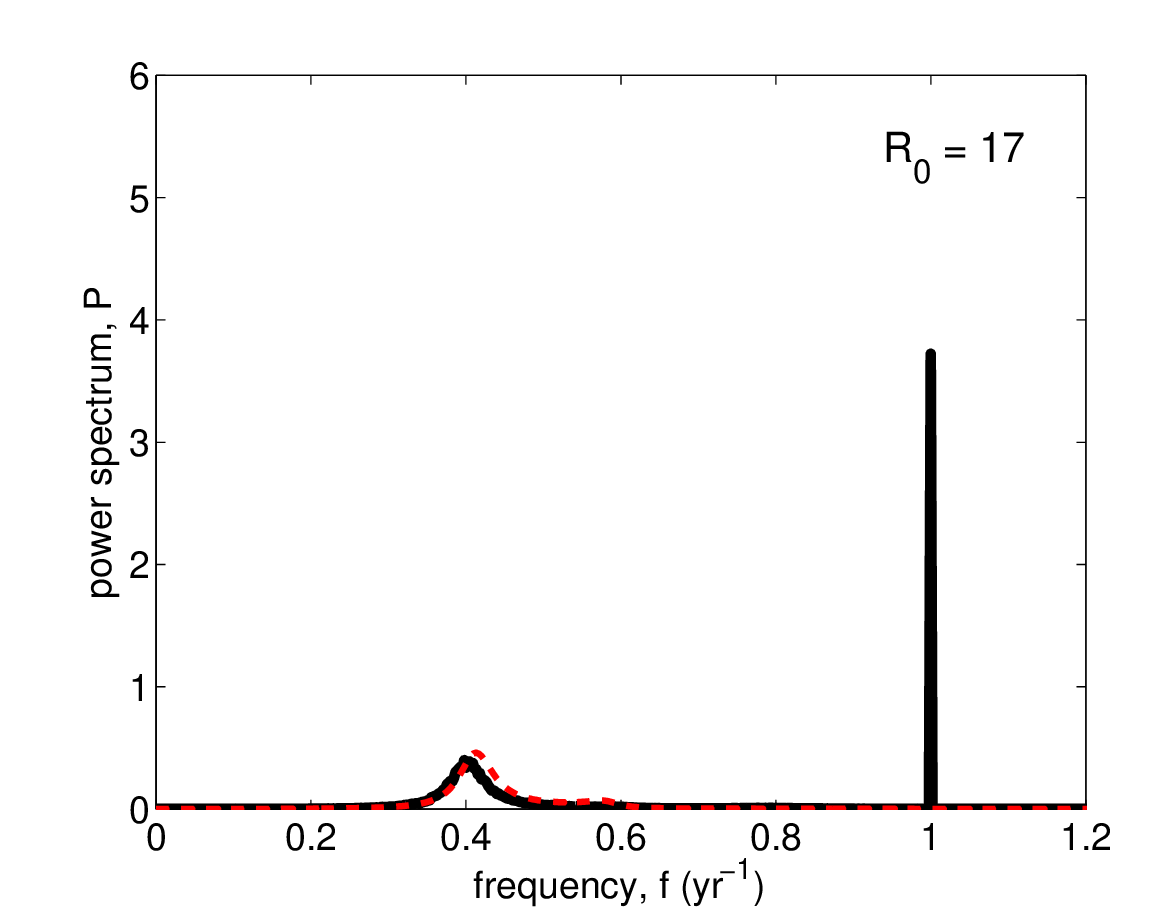}
\includegraphics[trim=0cm 0cm 0cm 0cm, clip=true, width=0.32\textwidth]{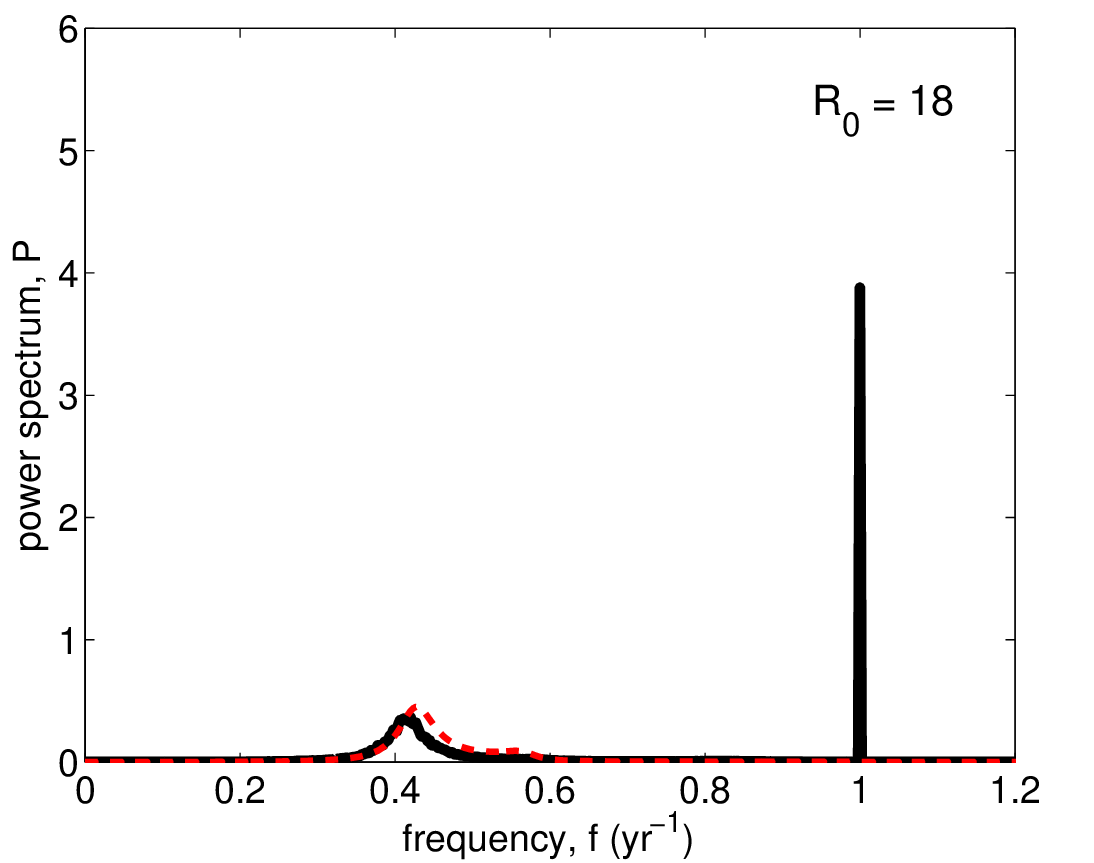}
{\caption{(Color online) Analytical spectra (dashed red lines) and full numerical spectra (solid black lines) for the seasonally forced model. Parameters: $\epsilon=0.05$, $\mu=0.02$ 1/y, $1/\nu=18$ d and $N=10^6$.}}
\label{fig2}
\end{figure}

\begin{figure*}
\centering
\includegraphics[trim=0cm 0cm 0cm 0cm, clip=true, width=0.48\textwidth]{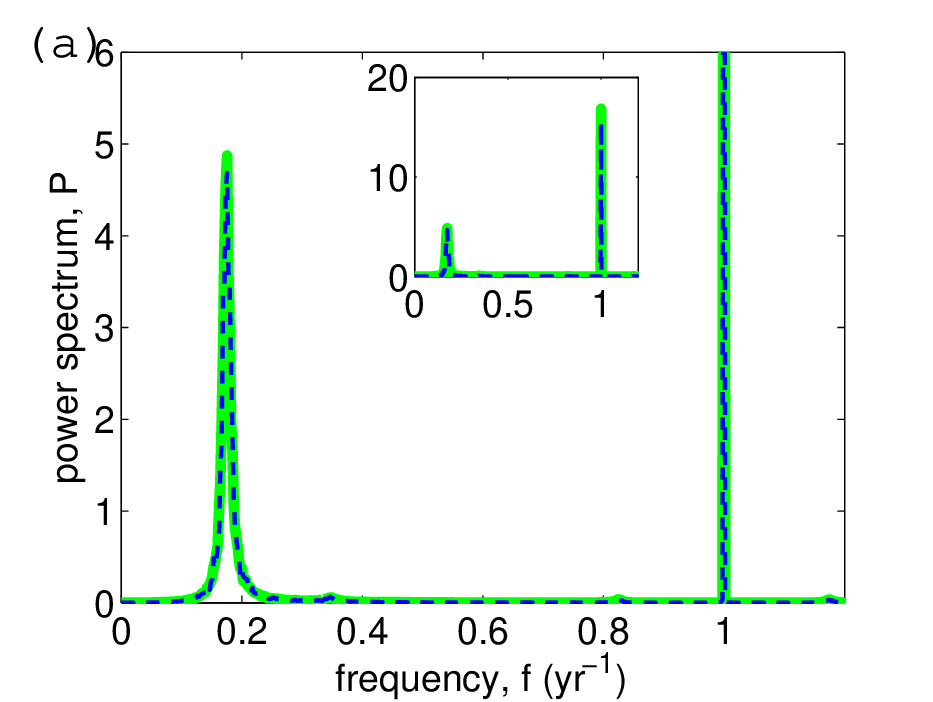}
\includegraphics[trim=0cm 0cm 0cm 0cm, clip=true, width=0.48\textwidth]{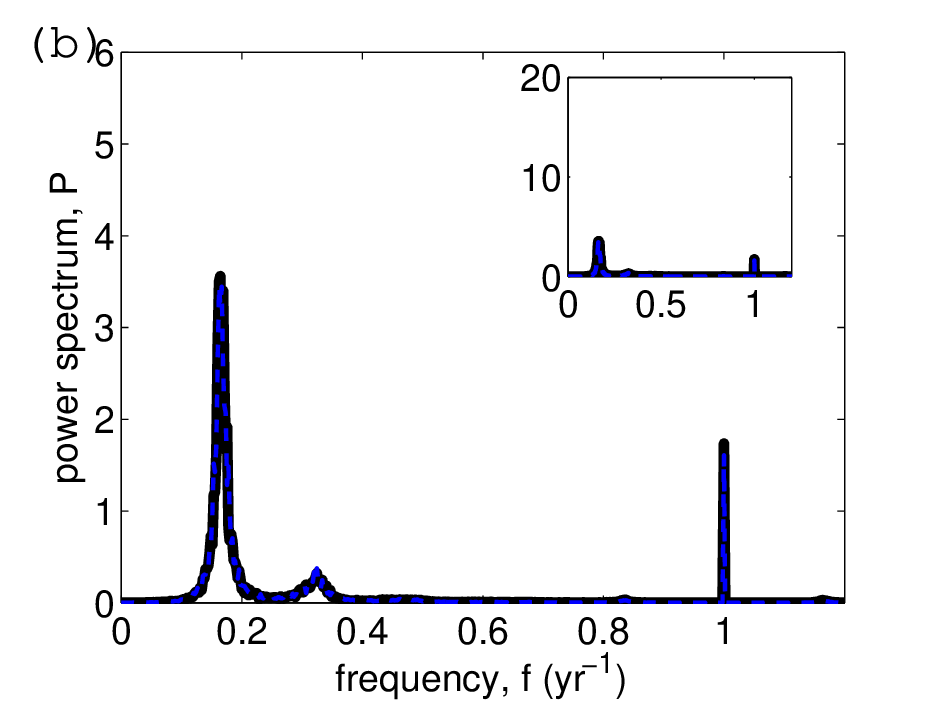}
{\caption{(Color online) Full numerical spectra for the model with sinusoidal forcing (solid green and black lines) and term-time forcing (dashed blue lines). The green and black lines are the same as in Figure 4 of the main text. In the term-time forced simulations we used $\beta(t)=\beta_0[1+\epsilon \text{Term}(t)]$, where $\text{Term}(t)$ is a periodic function that takes on value 1 during school terms and -1 during school holidays \cite{keeling:01,blackb:10}. The times of terms and holidays are taken from \cite{keeling:01}. These define the fraction of time spent in school, $f=92/365$. The basic reproductive ratio is given by $R_0\approx\langle\beta\rangle/(\nu+\mu)$ \cite{anderson:1991,keeling:01,blackb:10}, where $\langle\beta\rangle=\beta_0[(1+\epsilon) f + (1-\epsilon)(1-f)]$ is the time-averaged transmission rate. The perfect agreement between the spectra is achieved if the term-time forcing amplitude is 2.7 times larger than the sinusoidal forcing amplitude. This is consistent with the discussion in Ref [13] of \cite{earn:00} and in the Appendix B of \cite{bauch:2003}. Parameters: $\mu=0.02$ 1/y, $R_0=4$, $1/\nu=18$ d, $\epsilon=0.05$ (sinusoidal forcing, green and black lines), $\epsilon=0.135$ (term-time forcing, blue lines) (a) $N=10^7$ and (b) $N=10^6$.}}
\label{fig3}
\end{figure*}

\begin{figure*}
\centering
\includegraphics[trim=0cm 0cm 0cm 0cm, clip=true, width=0.48\textwidth]{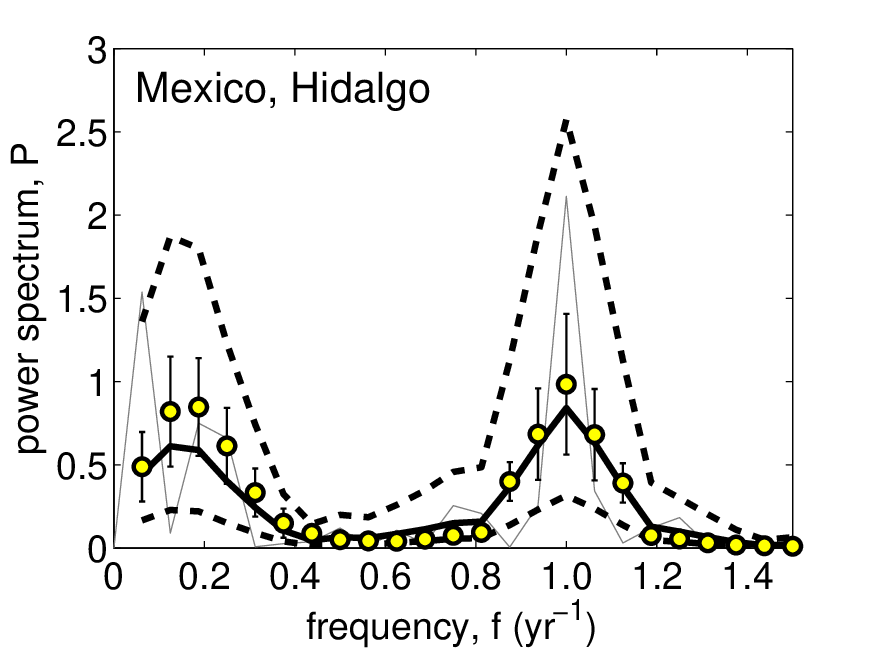}
{\caption{(Color online) Comparison of the full numerical spectrum of monthly incidence from simulations of the SIR model with seasonal forcing (yellow circles) and the spectrum obtained from the time series for Hidalgo, Mexico (lines). The lines are repeated from Figure 1 (d) of the main text. The simulated time series were 16 years long and sampled every month. The error bars correspond to the standard deviation for 20 simulations. In simulations we took $N=2100000$, the reporting rate of 0.01 \cite{metcalf:2011} and $R_0=4$. This value of $R_0$ is slightly lower than that estimated in \cite{metcalf:2011}. Its justification is two-fold: 1) the last years of the data time series correspond to post-vaccination and were not included in the analysis in \cite{metcalf:2011}; 2) the lifespan, $1/\mu$, is likely to be longer than the one we used for all data sets in the main text because the Mexico data is more recent. Parameters: $\mu=0.02$ 1/y, $1/\nu=18$ d, $\epsilon=0.17$.}}
\label{fig4}
\end{figure*}